\documentclass{elsart}
\usepackage{epsfig}
\usepackage{rotate}
\begin{document}

\setcounter{tocdepth}{1} 

\newcommand{\vo}{$V^0$}
\newcommand{\lam}{\mbox{$\rm \Lambda$}}
\newcommand{\alam}{\mbox{$\rm \overline{\Lambda}$}}
\newcommand{\ko}{\mbox{$\rm K^0_s$}}
\newcommand{\xizero}{\mbox{$\rm \Xi^0$}}
\newcommand{\pip}{\mbox{$\pi^+$}}
\newcommand{\pim}{\mbox{$\pi^-$}}
\newcommand{\gam}{\mbox{$\rm \gamma$}}
\newcommand{\al}{\mbox{$\rm \alpha$}}

\newcommand{\Fixup}[1]{{\bf <<< #1 !!! }}

\begin{frontmatter}

\title{The Drift Chambers Of The Nomad Experiment}

\author[DAPNIA]{M. Anfreville}
\author[Paris]{P. Astier}
\author[DAPNIA]{M. Authier}
\author[DAPNIA]{A. Baldisseri}
\author[Paris]{M. Banner}
\author[DAPNIA]{N. Besson}
\author[DAPNIA]{J. Bouchez}
\author[Paris]{A. Castera}
\author[DAPNIA]{O. Clou\'e}
\author[Paris]{J. Dumarchez}
\author[CERN]{L. Dumps}
\author[Paris]{E. Gangler}
\author[DAPNIA]{J. Gosset}
\author[DAPNIA]{C. Hagner}
\author[DAPNIA]{C. Jollec}
\author[Paris]{C. Lachaud}
\author[Paris]{A. Letessier-Selvon}
\author[Paris]{J.-M. Levy}
\author[CERN]{L. Linssen}
\author[DAPNIA]{J.-P. Meyer}
\author[DAPNIA]{J.-P. Ouriet}
\author[DAPNIA]{J.-P. Pass\'erieux}
\author[DAPNIA]{T. Pedrol Margaley}
\author[CERN]{A. Placci}
\author[DAPNIA]{A. Pluquet}
\author[DAPNIA]{J. Poinsignon}
\author[Paris]{B. A. Popov\thanksref{Dubna}}
\author[DAPNIA]{P. Rathouit}
\author[Paris]{K. Schahmaneche}
\author[DAPNIA]{T. Stolarczyk}
\author[Paris]{V. Uros}
\author[Paris]{F. Vannucci}
\author[DAPNIA]{M. K. Vo}
\author[DAPNIA]{H. Zaccone}

\address[CERN]{CERN, Geneva, Switzerland}
\address[DAPNIA]{DAPNIA, CEA Saclay, 91191
Gif-sur-Yvette Cedex, France}
\address[Paris]{LPNHE, Laboratoire de Physique Nucl\'eaire et 
des Hautes Energies,\\
Universit\'es de Paris 6 et 7, 75252 Paris Cedex 05, France}
\thanks[Dubna]{on leave from the Laboratory of Nuclear Problems, JINR, 
141980 Dubna, Russia}
\date{\today}

\begin{abstract}
We present a detailed description of the drift chambers 
used as an active target and a tracking device in the NOMAD experiment
at CERN.
The main characteristics of these chambers are a large area 
($\rm 3 \times 3~m^{2}$), a self 
supporting structure made of light composite materials and a low cost.
A spatial resolution of $\rm 150~\mu m$ has been achieved with a single hit
efficiency of $97\%$ .
\end{abstract}
\end{frontmatter}

\newpage
\section*{Contents}
\begin{tabular}{l l r}
1 & Introduction & 2 \\
2 & The drift chamber layout & 3 \\
3 & The drift chamber electronics & 10 \\
4 & The drift chamber construction & 13 \\
5 & Gas system and slow control & 17 \\
6 & The drift chamber performances & 18 \\
7 & The drift chamber reconstruction software & 25 \\
8 & Check of the drift chamber performances using experimental data & 35 \\
9 & Conclusions & 40 \\
& Acknowledgements & 40 \\
& References & 40 \\
\end{tabular}
%
%
\section{Introduction}
\label{sec:introduction}
The NOMAD experiment~\cite{NOMAD_Proposal} was 
built to search for  $\rm \nu_\mu \rightarrow \nu_\tau$ oscillations
in the CERN SPS neutrino beam predominantly composed of 
$\rm \nu_\mu$'s with a mean energy of 24 GeV. 
The search~\cite{NOMAD_OSC} was based on the identification of
$\rm \tau$'s produced by $\rm \nu_\tau$'s charged current (CC) interactions: 
$\rm \nu_\tau + N \rightarrow \tau^- + X$. \\
Given the $\rm \tau$ lifetime and the average energy of the CERN SPS 
neutrino beam, $\rm \tau$'s travel about 1 $\rm mm$ before decaying.\\ 
The spatial resolution of the NOMAD detector, though good, is not
sufficient to resolve such short tracks. Instead, the decaying 
$\rm \tau$'s are identified through the kinematics of their decay products. 
The presence of at least one neutrino in   
the final state allows using momentum balance in the plane
perpendicular to the neutrino beam direction in order to select 
$\rm \nu_\tau$ CC interaction candidates in a
copious $\rm \stackrel{(-)}{\nu}_{e, \mu}$ CC and NC 
background~\cite{nutau_idea}.
By studying correlations between the sizes and directions of three vectors
in the transverse plane
($\rm \vec p_T^{\ miss}$, $\rm \vec p_T^{\ lepton}$ and $\rm \vec p_T^{\ hadrons}$)
one can distinguish events containing $\tau^-$ decays from different sources
of background, provided that the event kinematics is well reconstructed. 
This method of oscillation search therefore 
necessitated an excellent quality of measurement 
of all the secondary particles produced in the neutrino interactions
and identification of electrons and muons. 
This was made possible thanks to an
active target located inside a dipole magnet~\cite{UA1}, 
and consisting of a set of large
drift chambers, providing at the same time the neutrino target material 
and the charged particles tracker. Given this dual role, the chambers had to meet two conflicting requirements:
their walls had to be as massive as possible in order to maximize the number of
neutrino interactions and as light as possible to limit multiple
scattering, secondary particle interactions and photon conversions. 
These conditions imposed the use of a low $Z$ material with good mechanical 
properties, mainly composite plastic materials.\\ 
The end result of the various necessary compromises was a target of 2.7 tons
fiducial mass over a total 
volume of $\rm 2.6 \times 2.6 \times 4.5~m^{3}$.
This gave a low average density of about 
0.1 $\rm g/cm^3$ and less than 1\% of a radiation length between two
consecutive measurements.

The NOMAD detector was built 
over a period of
4 years starting in 1991 and was
in operation during 
the next
4 years from 1995 to 1998.
This paper is devoted to a description of the drift chambers used in this 
experiment.

%
%
\section{The drift chamber layout}
\label{sec:mecanique}
\subsection{Introduction}  
The use of drift chambers for large detection areas 
appears to be the
best solution for the reconstruction of charged particle trajectories.
The event rate in neutrino experiments is such that no serious problem of 
pile-up can occur.
The main requirement for the physics studied in NOMAD was
to have a high density of coordinate measurements with good spatial resolution
in order to apply the kinematic selection method described above, as well as
to obtain a good determination of the neutrino interaction point 
and to be able to distinguish an
electron from $\rm \tau$ decay from a converted gamma ray. 
The traditional solution is to weave wires over a rigid frame in order to produce the
drift field. 
In our case the overall dimensions were limited by the internal size of the available magnet.
Rigid frames would have reduced further the active area.
The retained solution uses
rigid panels of composite material on which glued aluminium strips
produce the shaping field in the drift cell. 
Each chamber consists of three measurement planes allowing 
the reconstruction of one space point.

\subsection{General overview}
\begin{figure}[h]                                                               
\begin{center}                                                               
   \mbox{                                                                    
     \epsfig{file=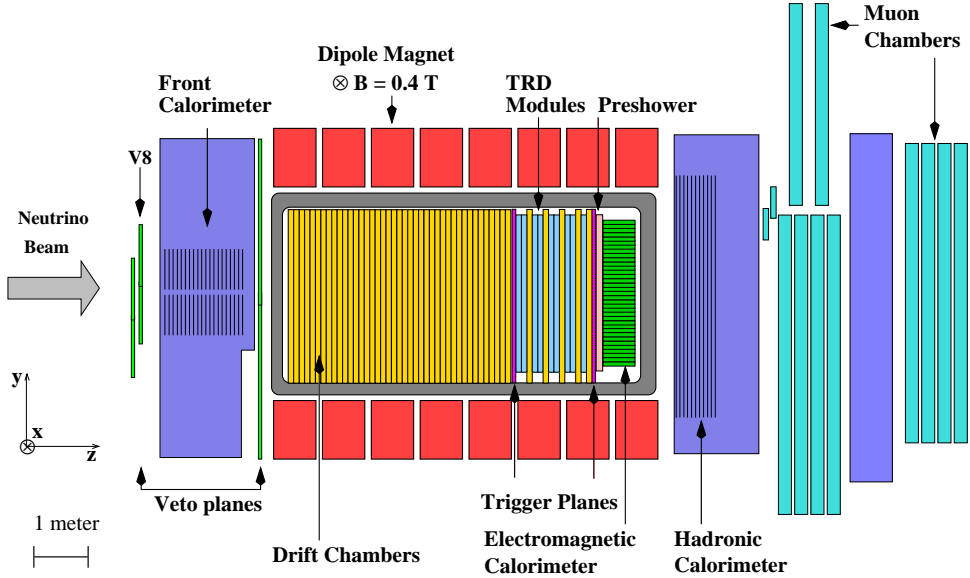,width=130mm}}
     \caption{A sideview of the NOMAD detector.
Z axis is horizontal and nearly coincides with the neutrino beam direction.
Y axis is vertical and points to the top of the figure. X axis is along the
magnetic field direction.}
      \label{fig:nomad_detector}
   \end{center}                                                                 
\end{figure}

The 
fiducial volume
of the NOMAD detector~\cite{NOMAD_Nimpaper} 
consists of 44 drift chambers which act both as a low-density target
and a tracking device. It is placed in a dipole magnet of internal
volume $\rm 7 \times 3.5 \times 3.5 ~ m^3$
operated at 0.4 Tesla. 
A sideview of the NOMAD detector is shown in figure~\ref{fig:nomad_detector}.
The magnetic field direction which is horizontal and orthogonal 
to the beam axis has been chosen as the X reference axis. The vertical axis is
called the Y reference axis and the Z axis is obtained by the vectorial product
of the X and Y directions (see figure \ref{fig:nomad_detector}).
Downstream of the 4 m long target, 5 extra drift chambers are inserted in between the modules of the transition
radiation detector (TRD)~\cite{NOMAD_TRD} in order to complete the tracking down to the 
preshower detector placed
in front of the electromagnetic calorimeter~\cite{NOMAD_ECAL}, and to improve the
momentum measurement resolution.

\begin{figure}[h]                                                               
  \begin{center}                                                               
    \mbox{                                                                    
    \epsfig{file=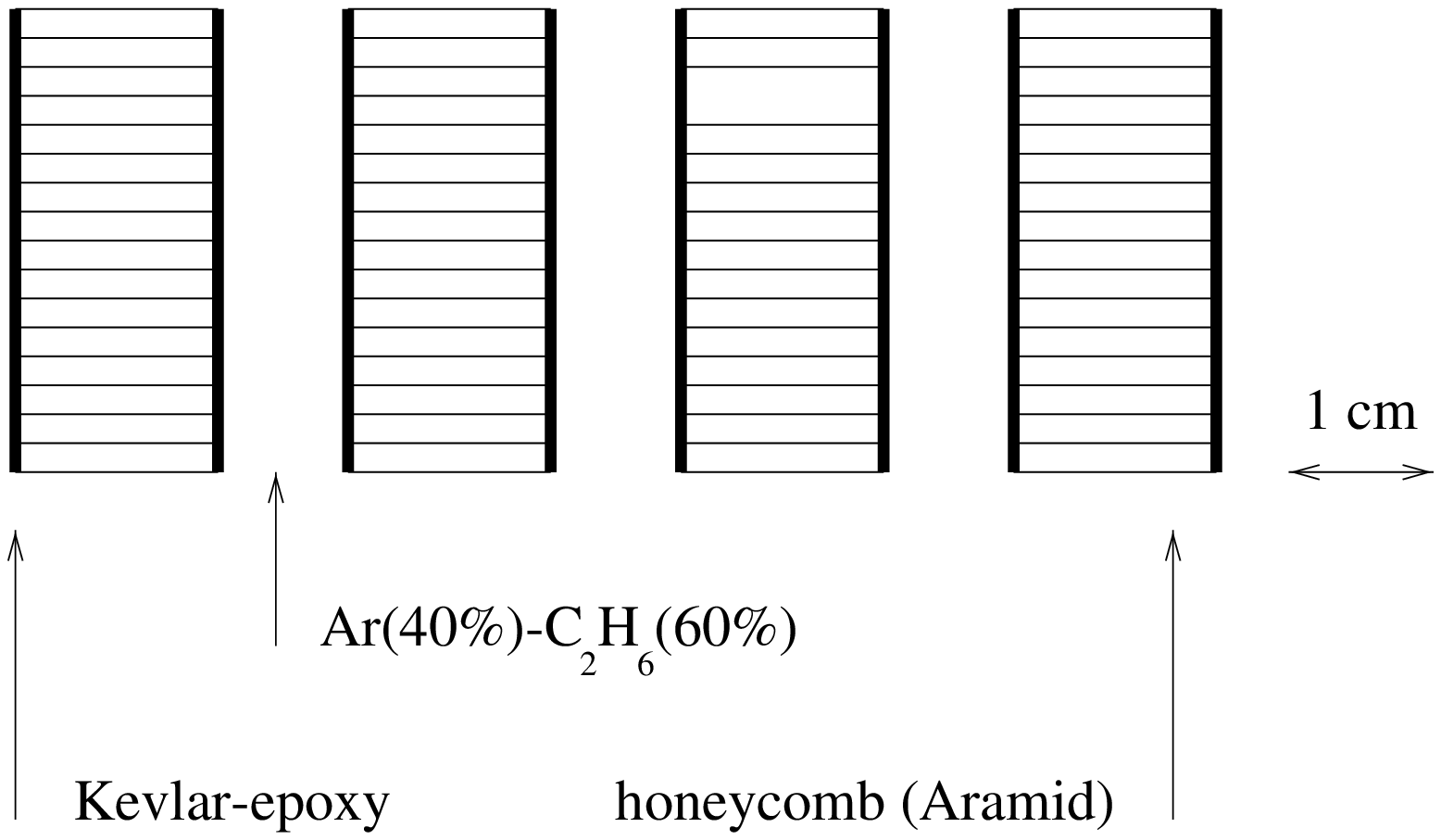,width=100mm}}
  \end{center}
  \caption{A sideview of one NOMAD drift chamber cut by a plane orthogonal
     to the X axis.}
  \label{fig:dc_cut}                                                                
\end{figure}

Each drift chamber has an active 
area
of about $\rm 3~m$ by $\rm 3~m$ and consists of four
panels enclosing 3 drift gaps of $\rm 8~mm$ 
(see figure~\ref{fig:dc_cut}) 
filled 
with an $\rm Ar(40\%)-C_2H_6(60\%)$
gas mixture at atmospheric pressure. The central gap (Y plane) is equipped 
with 44 sense
wires parallel to the X axis and the outer gaps have 41 wires at $\rm - 5^{\circ}$ 
(U plane) 
and $\rm +5^{\circ}$ (V plane) with respect to the X axis. 
For a high momentum track
crossing the chambers along the Z axis the Y coordinate is obtained from the Y
plane whereas the X coordinate is calculated combining the U and V plane
measurements.
Because of the small stereo angle, the resolution in X is about 10
times worse than the one in Y.

We start from a mechanical description of the drift chambers, while 
a detailed description of the drift cell will be given in 
subsection~\ref{subsec:cell} (figure~\ref{fig:ht_cell}).

\subsection{The panels} \label{sec:panels}

The design was studied in order to get self supporting chambers which act as
a neutrino target. The challenge was to obtain a rigid and flat
surface of $\rm 3 \times 3~m^{2}$ which is at the same time as ``transparent'' as 
possible to particles and massive enough to
yield a significant number of neutrino interactions.
For these reasons, the panels have a composite
sandwich structure of low Z materials. Each of them is composed of two $\rm 0.5~mm$
kevlar-epoxy ($\rm 0.57~kg/m^2$) skins surrounding an aramid honeycomb core structure
($\rm 32~kg/m^3$, 15 $\rm mm$ thick). 
Other solutions like polystyrene skins with rohacell or polystyrene foam have been tested and
excluded 
because of 
rigidity and flatness considerations,
although 
both
solutions
worked for small area (less than 3~$\rm m^2$) prototypes.

The total amount of material in each panel corresponds
to 0.5\% of a radiation length. 
The total thickness of a chamber is about $\rm 10~cm$ in Z and each chamber 
contributes 2\% of a radiation length. The total fiducial mass (including the glue and the strips) of the
44 chambers is 2.7 tons over an area of $\rm 2.6 \times 2.6~m^{2}$. The target is
nearly
isoscalar
($\rm N_{protons} \ : \ N_{neutrons} \ = \ 52.4\% \ : \ 47.6\%$).
The material composition is shown in table \ref{TABLE_MAT_COMPOSITION}.
\begin{table}[htb]
\begin{center}
\begin{tabular}{|l|c|c|c|}
\hline
Atom &prop./weight (\%) &protons (\%) & neutrons (\%) \\
\hline
C  & 64.30 & 32.12 & 32.18 \\
H  & 5.14  & 5.09  & 0.05  \\
O  & 22.13 & 11.07 & 11.07 \\
N  & 5.92  & 2.96  & 2.96  \\
Cl & 0.30  & 0.14  & 0.16  \\
Al & 1.71  & 0.82  & 0.89  \\
Si & 0.27  & 0.13  & 0.14  \\
Ar & 0.19  & 0.09  & 0.10  \\
Cu & 0.03  & 0.01  & 0.02  \\
\hline
Total & 99.99 & 52.43 & 47.56 \\
\hline
\end{tabular}
\caption {\label{TABLE_MAT_COMPOSITION}{Drift chamber composition by weight (in percent).}}
\end{center}
\end{table}

The panel structure is reinforced by replacing locally the honeycomb
with melamine inserts in the center and at
the corners. 9 spacers ($\rm 10~mm$ in diameter) placed 
between two panels
maintain the $\rm 8~mm$ drift gap. Nine 
insulated
screws, 
$\rm 2~mm$ in diameter,
cross
the chambers in the center of the spacers in
order to reinforce 
the whole structure. 
Four  $\rm 10~mm$ diameter 
screws go through the whole chamber (one at each corner) and are used 
both for chamber 
assembly and panel positioning.

\subsection{The drift field strips}
\begin{figure}[h]                                                               
  \begin{center}                                                               
    \mbox{                                                                    
    \epsfig{file=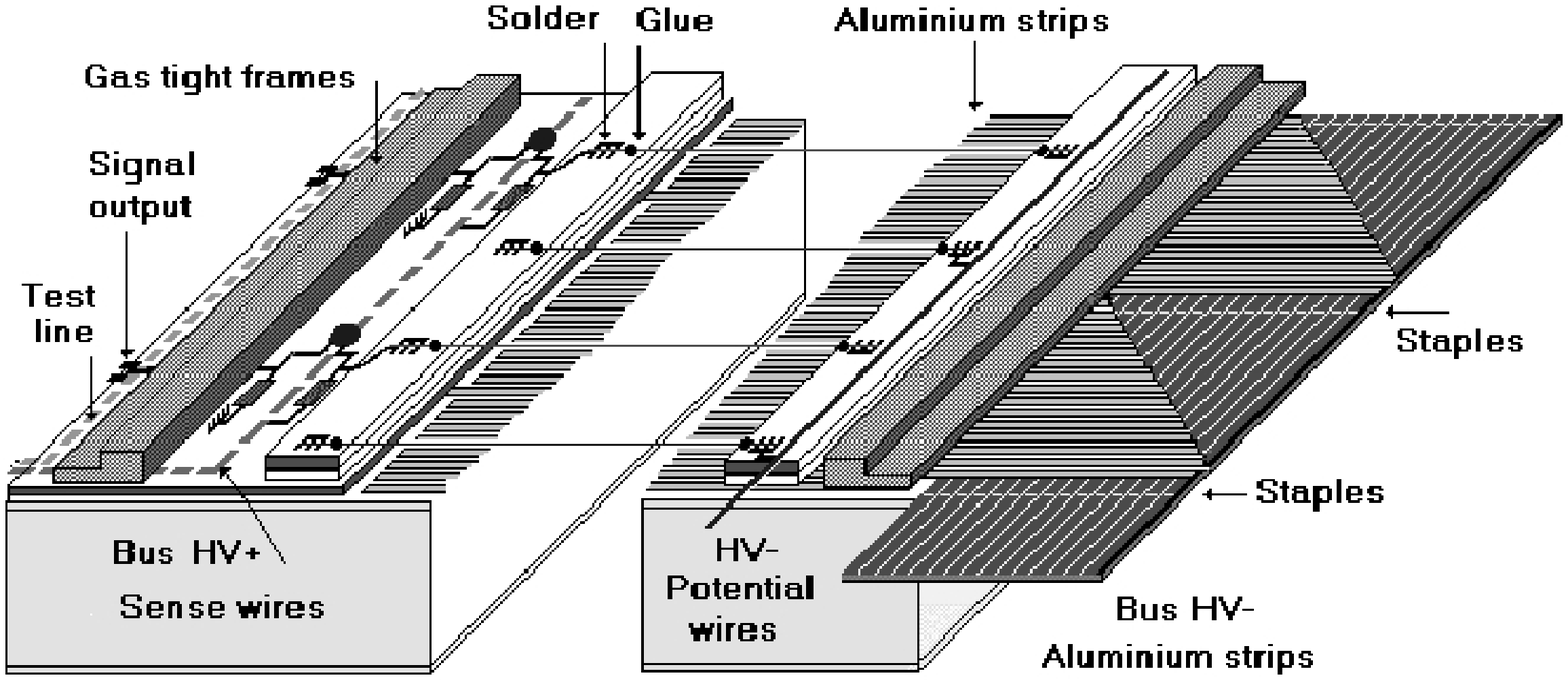,width=150mm,height=5.5cm}}
    \end{center}
  \caption{
A cut view of a panel equipped as a Y plane.
}
  \label{fig:dc_connections}                                                                
\end{figure}

To form an appropriate electric field in the drift cells,
a polyester skin with aluminium strips is glued  
on each side of a drift gap. 
The strips are $\rm 2.8~mm$ wide, $\rm 12~\mu m$ thick and separated
by a $\rm 1.2~mm$ gap large enough to avoid any sparking due to
potential differences between strips. The sparking potential difference is 2200 V 
whereas the applied potential difference between adjacent strips is 400 V.
The strips are obtained by a serigraphic
technique. Ink is deposited on an aluminized polyester band 
with a 
144 strip pattern. The band is then chemically treated in order
to remove the aluminium between the strips. Finally the ink is washed off.
The 5 strip bands needed to
cover the full plane are tilted as the wires, 
with respect to the X axis by $\rm -5^{\circ}$, $\rm 0^{\circ}$ and
$\rm +5^{\circ}$ depending on the drift gap. A gas-tight frame of bakelized paper, $\rm 4~mm$ 
thick, is glued on the panel to close the drift gap. At one end the 
strips go beyond the frame and the 16 strips corresponding to each 
drift cell (see subsection \ref{subsec:cell}) are connected to a flexible high 
voltage bus 
using
staples (see figure \ref{fig:dc_connections}). The 6 HV buses 
of 
each
chamber are connected to a high voltage distribution board located at 
the bottom of the chamber (see figure \ref{fig:ht_cell}). 
Each pair of HV buses corresponding to a drift plane
has its own resistor chain connected to a HV power supply. 
For the U and V planes ($\rm -5^{\circ}$ and $\rm +5^{\circ}$) a triangular zone of 
$\rm 25~cm$ by $\rm 3~m$ is
left unequipped and is covered with glue in order to avoid any gas leaks through
the panel. The residual distribution of the measured strip
band positions with respect to the design values has an RMS of 
$\rm 250~\mu m$ and
the positioning accuracy has been kept 
better\footnote{All the strip bands with a positioning error greater than 
$\rm 500~\mu m$ have been replaced.}
than $\rm 500~\mu m$.
We tried two methods for gluing of the strip bands. The first one was to use adhesive techniques because of easy
manipulation during the construction. We tested the stability with time of 
this gluing technique in a
vessel filled with an $\rm Ar(40\%)-C_2H_6(60\%)$ gas mixture during more than
2 years. This test was also used to check the aging of different materials
(polystyrene, kevlar, rohacell, etc.). No change was observed.
However, chambers built with this technique suffered from short circuits after several weeks
of operation. Opening the chambers, we saw gas bubbles between the panels and the
strip bands. This is probably due to the moisture gradient
between the outside of the chamber and the dry gas in the drift gap.
About 25 chambers suffered from the problem and they had to be modified 
both at Saclay and in a special workshop set up at CERN.
The final gluing technique used a bi-component polyurethane glue to
provide a stronger and more reliable 
attachment
of the strip bands.

\subsection{The drift cells}\label{subsec:cell}
\begin{figure}[h]                                                     
  \begin{center}							       
    \mbox{								      
    \epsfig{file=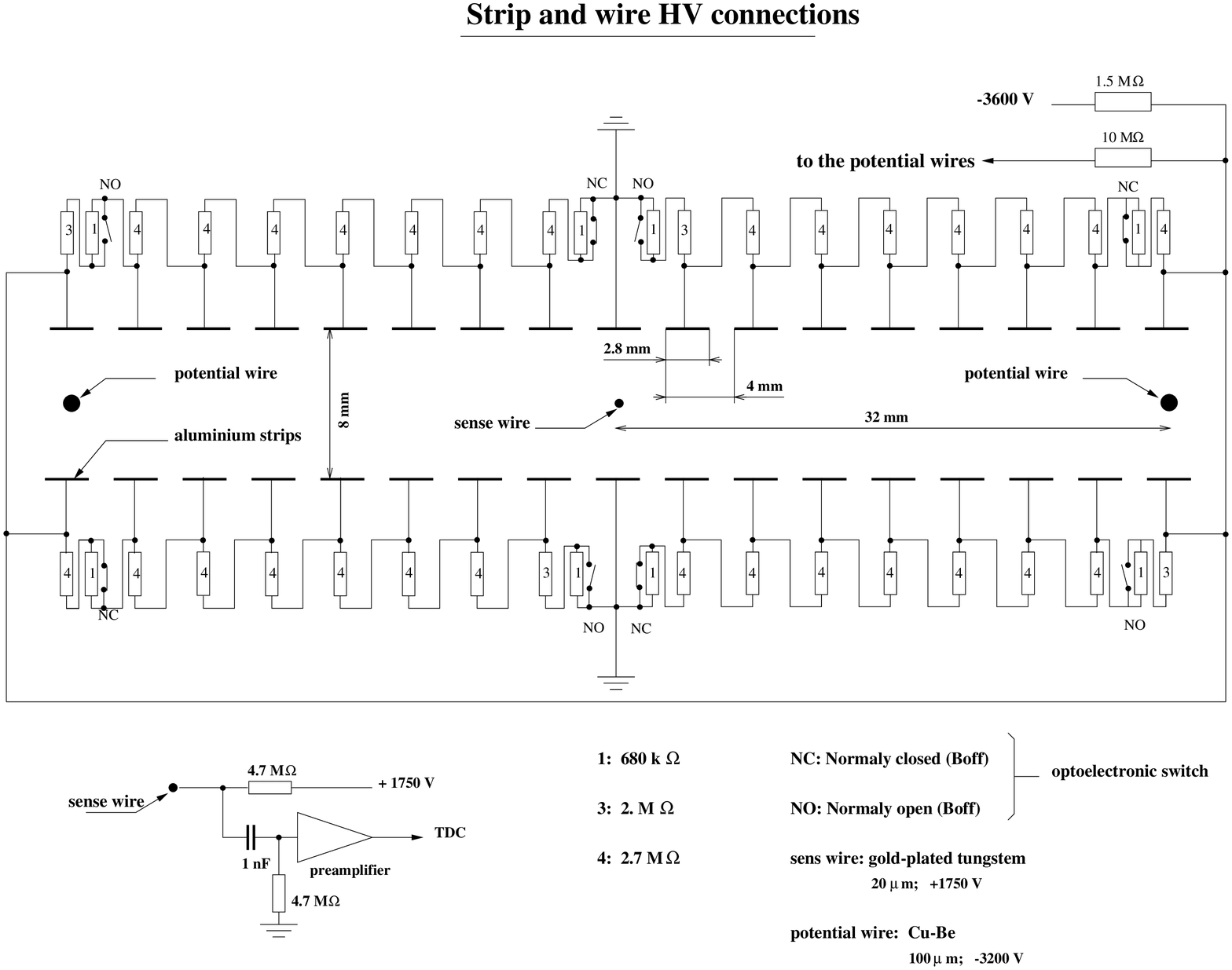,width=\linewidth}}
  \end{center}
  \caption{A close-up of a drift cell and its HV connections.}
  \label{fig:ht_cell}
\end{figure}

The $\rm 20~\mu m$ diameter gold-plated tungsten sense wires placed in the center
of the drift cells are held at a typical voltage of $\rm +1750~V$ 
through a $\rm 4.7~M\Omega$ resistor 
(see figure \ref{fig:ht_cell}). 
The corresponding gain is of the order of $10^5$.
The sense wire signal is readout by a 
preamplifier located on a board connected directly to the chamber printed board 
and AC coupled through a $\rm 1~nF$ capacitor. 
The preamplifier signal goes through a discriminator and is sent via a $\rm 30~m$
long cable to the TDC's (see section \ref{sec:electronique}).
A test line located on the chamber printed board outside of
the gas tight frames (see figure
\ref{fig:dc_connections}) allows to inject a signal just 
at the entrance of
the amplifiers for test purposes.

Two Cu-Be 
potential wires, $\rm 100~\mu m$ in diameter, 
are placed at a distance of $\rm \pm 32~mm$ with respect to the sense
wire and held at $\rm -3200~V$ (see figure \ref{fig:ht_cell}) through a 
$\rm 10~M\Omega$ resistor 
which is 
shared by
all potential wires of a drift plane.

\begin{figure}[h]                                                                                                                
  \begin{center}							       
    \mbox{								      
    \epsfig{file=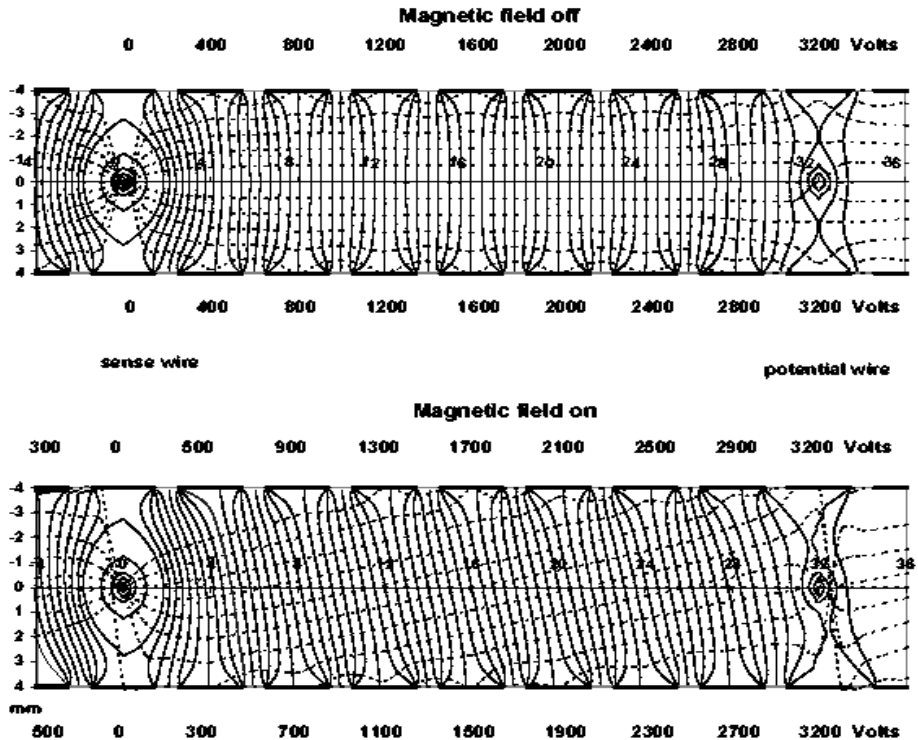,width=110mm,height=13cm,angle=-90}}
  \end{center} 
  \caption{
Potential and field lines along the drift cell
for the two modes of operation: 
field
OFF (top) and 
field
ON (bottom).}
  \label{fig:dc_BonBoff}					   
\end{figure}

The strips directly in front of the sense wires are grounded whereas those 
in front of the potential wires are held at $\rm -3200~V$. The strips in-between 
have a potential equally distributed so that the drift field
perpendicular to the sense wire is of the order of $\rm 1~kV/cm$ in most of the cell (see figure \ref{fig:dc_BonBoff}).
The measured drift velocity is approximately
$\rm V_{drift} = 50~\mu m/ns$ as expected from
\cite{Sauli_drift}. However, close to the sense and potential wires, the drift field
changes 
drastically
introducing non-linearities in the time to distance relation
which are taken into account as second order corrections. 

Since the drift chambers have to be operated inside a magnet delivering a
field of 0.4 Tesla, the drift direction is tilted due to the Lorentz force.
This effect 
is
compensated for 
by shifting the
potential on opposite strips by $\rm \pm 100$ V 
\cite{Charpak_tilt1,Charpak_tilt2} 
(see figure~\ref{fig:dc_BonBoff}). 
Field configurations for 
field
on and 
field off operation modes 
are obtained through 
an optoelectronic 
switch
located on the HV distribution board (see figure \ref{fig:ht_cell}).

The 3 $\rm m$ long 
potential
and sense wires are soldered and glued at both ends 
(see figure~\ref{fig:dc_connections}) with a precision of about $\rm 100~\mu
m$. Since the time to distance relation is 
extremely sensitive to
the relative position of
the wires with respect to the strips, the dominant error comes from the strip
positioning. The final knowledge of the wire position in the
experiment will be the result of a software alignment procedure described in
section~\ref{sec:alignement}: in order to reduce the gravitational and 
electrostatic sagitta\footnote{From our
prototype studies, this proved to be a necessity to reach the 
desired spatial resolution~\cite{Victor}.},
the wires are glued on 3 epoxy-glass rods of 5 $\rm mm$ width which are 
in turn
glued (parallel to the Y axis)
over the strips. The free wire length is therefore reduced by a factor of 4 at the price of
three dead regions, each less than $\rm 1~cm$ 
wide. 
To avoid having the dead regions aligned
in the detector, the positions of the support rods are staggered along X by $\rm -5~cm$ (U plane), 
$\rm 0~cm$ (Y plane) and $\rm +5~cm$ (V plane).

\subsection{The gas supply}
The argon-ethane mixture is provided from the bottom of each chamber at 3 millibars
above atmospheric pressure by two plastic pipes (one at each chamber side). The
gas enters a $\rm 20~mm$ diameter distribution pipe which goes across the
chamber. At each drift plane, the gas diffuses through $\rm 5~mm$ holes into the gap.
On top of the chamber a similar system collects the gas. 
The gas tightness of each drift gap is ensured by inserting a
string of polymerized silicon joint between the 
two 
frames.
A second method using clamps and elastic o-ring joints has been tested
and also used. The 
gas leaks 
were slightly
higher with this method.

\subsection{The chambers and modules}

In each drift cell, the hit position ($\rm d_h$) with respect to the wire is 
obtained using the measured drift time ($\rm t_d$)
and the time to distance relation ($\rm d_h = V_{drift} \times t_d$ in first
approximation). However, we do not know if the track has crossed the cell 
above
or below the sense wire (up-down ambiguity). In order to reduce this
ambiguity we have built two types of chambers differing only by the position
of the drift wires along the Y axis. The so-called Up chambers are moved upward by
$\rm +32~mm$ in Y (half a cell) with respect to the Down chambers.
In the target region, the chambers are grouped in modules of 4 chambers 
with a pattern corresponding to Up, Down, Down, Up, thus avoiding 
aligned sense wires. In the TRD region the distance
between two chambers is large (about $\rm 25~cm$); they are alternatively of type Up
and Down. The chambers grouped in modules are held together by 
6 stainless steel 
screws, $\rm 16~mm$ in diameter, crossing the chambers (3 at the top and 3 at 
the bottom).
The modules and individual chambers are supported by rails fixed
on the top of the overall central detector support, the so-called basket 
\cite{NOMAD_Nimpaper}.

%
%
\section{The drift chamber electronics}
\label{sec:electronique}
\subsection{The preamplifier} \label{sec:preamplifier}
\begin{figure}[h]
\begin{center} 
\mbox{ 
\epsfig{file=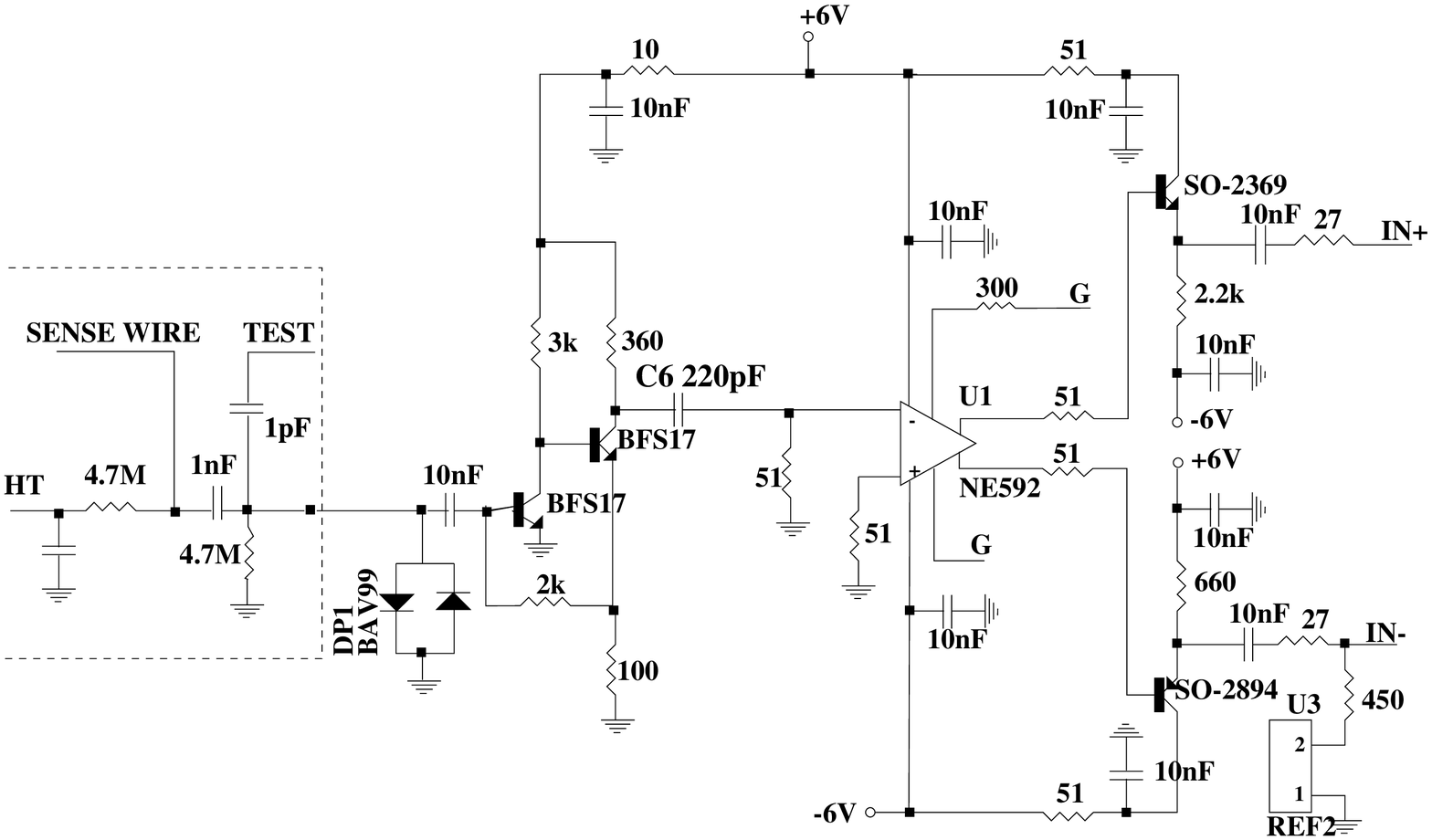,width=125mm}}
\caption{Schematic diagram of the preamplifier.}\label{preamp}
\end{center}
\end{figure}

The signal generated by an avalanche on the sense wire can be considered as a 
current generator. Therefore a transimpedance amplifier, 
connected to one end of the wire through a capacitor, is well adapted to our problem. 
The 
$20~\mu$m diameter 
gold-plated tungsten wire has a resistance of $165~\mathrm{\Omega}$ 
per meter. The characteristic impedance is not real and for high frequencies
(larger than $100~\mathrm{MHz}$) has a value of $360~\mathrm{\Omega}$. It is therefore difficult to 
match the line in order to avoid reflections, which are 
small due to 
the large attenuation of such lines. After several tests, the best solution 
turned out to leave the other end of the wire unconnected.
The schematic of the preamplifier is shown in figure~\ref{preamp}.
The input impedance is always 
small compared to the line impedance at high frequency. 

The amplifier is linear in the range 0 to $120~\mathrm{\mu A}$ input current. 
The high 
frequency cut off is $200~\mathrm{MHz}$. In order to improve the double pulse separation, the C6 capacitor (see figure~\ref{preamp}) 
was set to  $220~\mathrm{pF}$. This value fixed  
the lower frequency cut off of the amplifier to $1.76~\mathrm{MHz}$. 
It corresponds to a 
derivative of the signal with a time constant of $90~\mathrm{ns}$.  
In this frequency range the input impedance is below $\rm 10 \ \Omega$.
The preamplifier drives a LeCroy MVL 
4075 comparator. The threshold of the comparator can be adjusted to correspond
 to input currents between 1 and $\rm 100~\mu$A. During data taking this threshold was 
 fixed at $10~\mu A$ in order to keep pickup noise at a minimum. 
The threshold corresponds to approximately 15 electrons (i.e. 5 primary pairs).
The comparator differential ECL output goes true whenever the input signal 
is above the chosen threshold. 
The output signal is sent through a $30~\mathrm{m}$ long twisted pair line to the 
 control room and drives a TDC described later.
 
\begin{figure}[h]
\begin{center}
\epsfig{file=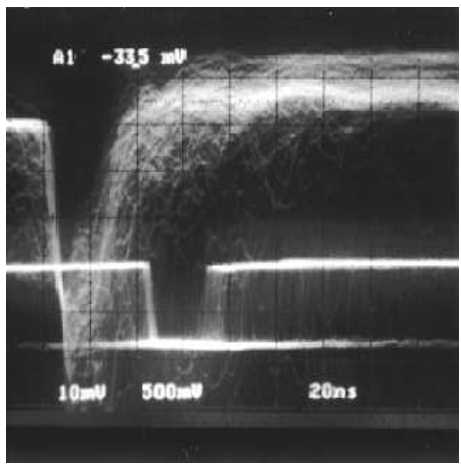,angle=90,width=60mm}
\caption{Chamber signal at the 
entrance of the comparator and the shaped signal.}\label{marfig3}
\end{center}
\end{figure}

All the options described previously have been tested with a ruthenium source 
which delivers electrons in the energy range up to $3.5~\mathrm{MeV}$,
able to 
cross one chamber panel easily. Figure~\ref{marfig3} shows the signal at the 
entrance of the comparator as well as the shaped signal. The small positive 
offset is the consequence of the derivative imposed on the signal. The 
price to pay for this better pulse separation capability is an 
increase of multiple ECL 
pulses when a particle crosses the chamber near the sense wire 
(see section \ref{subsec:afterpulses}).

The $3~\mathrm{m}$ long sense wires are very efficient antennas and collect easily radio 
emitters and noise. The drift cell field shaping strips provide already a good 
shielding which has been reinforced by a complete shielding of the chamber 
with a $12~\mathrm{\mu m}$ thick aluminium 
foil.
Tests have shown that thicker shielding 
does not improve the 
situation.
The chamber is essentially composed of 
insulating
material. It is therefore necessary to define properly ground 
references. For this purpose, the amplifiers have been mounted on a board 
whose ground support covers the entire board. On the side of the 
chamber panels, $3~\mathrm{mm}$ thick copper bars have been implanted and the ground 
plane of each electronic board is connected to them. We found experimentally 
that the strips in front of the sense wires had to be connected directly to the 
electronic ground of the 
board. Although the strip was already connected to ground through the voltage 
divider, this operation reduces the radio and noise background by more than 
one order of magnitude.

\subsection{The TDC's} \label{sec:tdc}

The signals from the discriminators are encoded by modified LeCroy multi-hit TDCs 1876
\cite{LeCroy} operated in a ``common stop'' mode. The TDCs hold in memory at most the last 16 hits for each channel
over a time period of $\rm 64~\mu s$.

To encode the signal on sixteen bits ($\rm 1~ns$ lsb corresponding approximately to $\rm 50~\mu m$ drift
distance), these TDCs :

\begin{enumerate}
\item record the arrival time \( t_{start} \) of the ``hits''\footnote{%
A hit can be defined as the leading or the trailing edge of the incoming signal,
or both. The latter method was used in the lab to study the comparator output signal~\cite{habilJPM}.
During data taking, only the leading edges were encoded.
}, i.e. pulses from the drift chamber comparators. A counter performs a fourteen bit coarse measurement
at twice the board clock frequency (\( 125\, \mathrm{MHz} \)). Three delay
lines give an additional 2 bit resolution.
\item record the arrival time \( t_{stop} \) of the stop signal, i.e. the pulse from the
experimental trigger, delayed to encompass the maximum drift time\footnote{%
The necessary common-stop delays were realized on dedicated fan-out cards inside
each FASTBUS crate. An input channel on each TDC was used to encode the undelayed
trigger pulse, allowing to correct for the delay discrepancies and long term variations.
}.
\item store the difference \( t_{stop}-t_{start} \) for each valid hit. Thus, small TDC values correspond to large drift times.
\end{enumerate}
The counter structure guarantees that the integral non-linearity of the encoded 
time difference depends only on
the main clock oscillator stability. The differential non-linearity \emph{on}
\emph{the individual input signals} is induced by the internal delay lines and
the clock interleaving. Each board and channel was individually qualified
in the lab using a digital delay as reference \cite{SRS}. The integral
non linearity was found to be negligible. The overall error on each channel was
smaller than \( 800\, \mathrm{ps} \).

The TDCs were modified to store all converted informations from successive events
in internal buffers during the neutrino spills ($4~\mathrm {ms}$), so the dead-time is limited to the time needed to handle the Common Stop ($1~\mathrm{\mu s}$) and to transfer the data from the encoding chip to the buffer ($150~\mathrm {ns}$ per hit). Thus the dead time \( T_{D} \)
induced by the readout of the drift chambers can be expressed as :
\[
T_{D}=1\, \mathrm {\mu s}+150\, \mathrm {ns}*maximum\, \#hits\, per\, module\, per\, event\]
 To 
reduce
the dead time and to insure that no buffer overflow occurs,
the signals from a given chamber were distributed across different TDC modules.
The minimal time between successive hits was found to be $9.5~\mathrm{ns}$ with a standard deviation of  $0.9~\mathrm{ns}$ across all TDCs. Extremum values were  $5~\mathrm{ns}$ and  $20~\mathrm{ns}$.

%
%
\section{The drift chamber construction}
\label{sec:bancdetest}
The goal was to build more than 50 chambers within one and a half year. 
The production line was split into 6 main phases with 6 corresponding main shops:
\begin{itemize}
\item{panel drilling}
\item{strip positioning and gluing}
\item{frame and printed circuit board gluing}
\item{strip connections}
\item{wire positioning and soldering}
\item{electric tests and assembly}
\end{itemize}
Each phase had to be carried out within 5 days and was operated by two technicians. 
A chamber needed about 5 weeks to be built. Finally the chamber was tested with cosmic particles.

\subsection{Panel selection and drilling}

After sand-papering and a visual control of the state of the surfaces, 
concavity and thickness were taken into account to select the panels. 

In more than 200 panels, 19 holes had to be drilled: 4 in order to assemble 
the chamber and for precise positioning measurements, 9 for the spacers and 6 
to build a module (as described in subsection~\ref{sec:panels}). 
A $\rm 9~m^2$ gauge built with 
the same materials as the panels (aramid honeycomb + kevlar epoxy skins) 
was used. On this gauge metallic inserts supported drilling barrels. 

\subsection{Strip gluing}

The straightness of the strip  bands was checked before gluing them on the panels with a 
bi-component polyurethane glue. The glue toxicity led us to build a special 
closed area around the stripping shop with fast air replacement. The operators 
had to wear breathing masks and appropriate gloves.
To glue with precision the strips on the panels, we used a table with a frame 
allowing precise relative indexing between the panels and the strip laying tool 
developed by our CERN collaborators. After gluing each band of 144 strips 
(5 bands per panel), a measurement of the strip positions was performed with a  
position gauge  attached to the frame  and parallel to the strips. 
In case of bad positioning, it was possible to quickly remove the band
from the panel. 
After gluing of the 5 bands, a strip position measurement with respect 
to the reference holes was performed again.  

\subsection{Frames and printed circuit boards}

A week later, gas tight frames as well as   
printed boards for high voltage and signal connections 
were glued on the strips and 
the panels 
(see figure \ref{fig:dc_connections})
using epoxy glue.

\subsection{Strip connections}

Before connecting the strips to the flexible high voltage bus, an electric 
test was performed to find short-circuits between strips. 
An automatic pneumatic stapler was used to establish 4000 connections per chamber. 
Thereafter, the 3 wire supporting epoxy-glass rods were glued perpendicularly over the strips to reduce 
the gravitational and electrostatic sagitta of the wires.
Finally, another electric test was performed to check the continuity and insulation 
of the connected strips and the bus.

\subsection{Wire positioning and soldering}

Using a position gauge, 
reference lines for wire positioning
were engraved on the printed circuits at 
the two ends of the chambers. Then the wires were brought into alignment with engraved 
lines using a video camera equipped with a zoom. The  tensions were adjusted to $\rm 50~$g for the sense wires 
(just below the limit of elastic deformation)
and $\rm 200~g$ for the field wires.
 After soldering and gluing the wires on 
printed circuits at both ends, tensions were checked by a 
resonance frequency measurement.
After possible replacement of some wires showing bad tension, 
all wires were glued on the 3 
perpendicular epoxy-glass rods. Then the position of 10 sense wires with 
respect to the 4 pins used to close the chamber
(one at each corner of the chamber)  
were measured using an optical ruler. 
These measurements were used later on as a starting point for the global 
geometrical alignment. 

\subsection{Electric test and assembly}

We checked that
the test lines running 
on the chamber side 
induced signals on all the sense wires by capacitive coupling 
(see figure \ref{fig:dc_connections}). 
Then some electrical tests were performed on each plane.
 
First, the insulation resistance between the strips was measured. 
Then a negative high voltage ($\rm -7000~V$) was applied to the strips and to the field wires. 
Measurements were performed on the field wires in order to check the leakage
current. In case of 
a large
leakage current cleaning of the board 
was necessary. 
After decreasing the HV to $\rm -3200~V$ (the nominal value) positive 
high voltage was applied to the sense wires to check 
the leakage current.
Then during about 2 hours, a sense wire cleaning was performed by applying a 
negative high voltage on the sense wires and no voltage on the strips and field wires. 
We observed a current decreasing with time on the sense wires. When no more current 
variation
was 
detected
we stopped the cleaning process.
After these tests, the chamber was assembled. The planes were glued together with a 
silicon joint laid down along the gas tight frames. After closing a gap, the
same electrical tests were done before closing the next one. Once the three 
gaps were closed, the nine 
insulated
screws corresponding to the nine spacers and the gas connectors 
were put into place. 
Three days later, the chamber was filled with argon in order to verify the gas tightness. 

\subsection{Cosmic test}

\begin{figure}[h]                                                               
  \begin{center}                                                               
    \mbox{                                                                    
    \epsfig{file=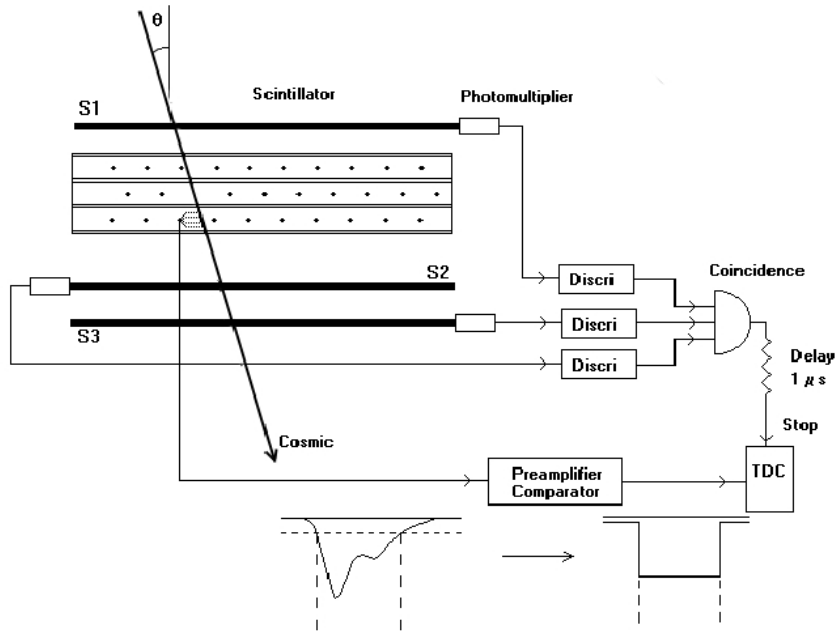,width=120mm}}
    \end{center}
  \caption{The cosmic test setup.}
  \label{fig:bancdetest_setup}                                                                
\end{figure}

\begin{figure}[h]                                                               
  \begin{center}                                                               
    \mbox{                                                                    
    \epsfig{file=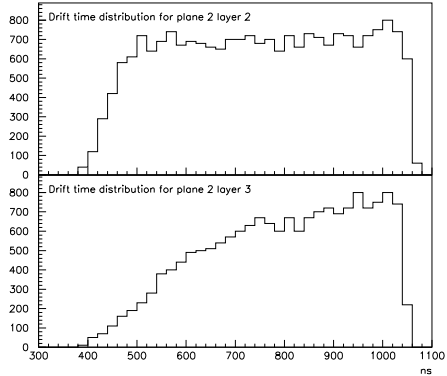,width=60mm}}
    \end{center}
  \caption{Drift time distribution as measured by the cosmic test setup. 
Top: for correctly aligned 
strip bands. Bottom:
  for misaligned ones.}
  \label{fig:bancdetest_derive}                                                              
\end{figure}

Before being sent to CERN for the final installation in the experimental area, 
each chamber was tested with cosmic rays~\cite{Mai}.
The experimental setup is shown in figure \ref{fig:bancdetest_setup}. 
The chamber was completely equipped with preamplifiers and filled with
$\rm Ar(40\%)-C_2H_6(60\%)$ gas mixture at atmospheric pressure. 
A first data taking was done by sending signals on the test lines with a pulser to 
check that the electronic chain (preamplifiers, cables, TDC) was functioning  correctly.
Then data taking with cosmic particles was performed. The coincidence of three 
scintillation counters, $\rm 3~m$ long and $\rm 30~cm$ wide, defined the crossing of a particle 
through the chamber.
The drift times and the number of hits on each wire were recorded. 
With the signal hit map it was possible to identify some potential problems: 
noisy wires, dead wires, efficiency losses, etc. 
We had to open some chambers to replace bad wires.
Drift time distributions could show some possible problems related to strip band positioning. 
Figure \ref{fig:bancdetest_derive} shows a non uniform drift time 
distribution observed with a bad strip 
band positioning following a mistake made during the gluing of the strips as compared to a normal drift
time distribution. 
The chamber 
was dismounted
and 
new strip bands 
were glued
on the panel.

%
%
\section{Gas system and slow control}
\label{sec:usineagazetslow}
In order to ensure a stable operation of the drift chambers during long
periods of time, a large number of parameters have to be kept under control,
such as the quality of the gas mixture, high and low voltages,
temperature, pressure, etc. 
A dedicated slow control system
has been developed to monitor these parameters and raise alarms when
necessary.

It was decided to use a common gas system for the target drift
chambers and the muon chambers of the NOMAD experiment.
The latter were muon chambers recycled from the UA1 experiment \cite{MUON},
with a total volume of 40 m$^3$ compared to 10 m$^3$ for the target
drift chambers, and both systems used a 40\% argon - 60\% ethane 
gas mixture. Such a mixture exhibits 
a large plateau in drift velocity 
as a function of the ratio 
of electric field over pressure, so that the effect of atmospheric
pressure variations is minimized.
The common input gas flow, whose composition was monitored every 10 mn,
was split between the two subsystems.
The gas was then distributed to each drift chamber, individual flows 
being adjusted using a manual valve. Cheap quarter turn 
valves were used and showed a good stability during the 4 years of
operation. A flowmeter on each chamber with digital reading 
(AWM3300 by Honeywell) allowed 
to perform the initial setup or subsequent adjustments of the valves.
Another flowmeter on the return circuit of each chamber allowed
to determine the output flow.
All these flowmeters had previously been 
individually calibrated both for the standard
argon-ethane mixture and for pure argon.
The flowmeters were interfaced with a MacIntosh running Labview~\cite{LABVIEW}
for a permanent monitoring of individual input and output gas flows.
\newline
Any significant departure from reference values for input or output flows
of individual chambers generated an alarm in the control room of the 
experiment.
The typical input flow for each chamber was set around 25 liters/hour
(the gas volume is 200 liters per chamber).
The typical leak rate was about 5 liters/hour. Some chambers had leaks
up to 12 liters/hour (in which case the input flow was increased to 30
liters/hour). Such high leak rates are inherent to the conception of these
chambers 
(due to the porosity of the panels),
but no loss in efficiency was observed for the chambers with 
higher leak rates. 
The oxygen content in the output gas mixture was about 500 ppm in the
global return circuit of all drift chambers and 
was continuously monitored to insure that there was no loss in efficiency.
A strong ventilation of the 
internal volume of the magnet ensured that no accumulation of gas from the
chambers could occur (safety gas detectors were installed in this volume). 
To avoid accidental overpressure, each chamber was protected by bubblers,
set up in parallel both on input and output circuits.
The gas flow through the drift chambers was assisted by an aspirator pump.
The gas from the return circuit was partly recycled after going through
oxygen and water traps, in order to save on ethane consumption.

The monitoring program was also controlling the correct value of the 
low voltages for the electronics of the chambers, 
including discriminator thresholds and
the setting of the 
$B/\bar{B}$ switch for strip high voltages 
(corresponding to the magnetic
field being on or off, see subsection~\ref{subsec:cell}); 
it also controlled the values of
the high
voltages (through a CAMAC interface to the CAEN power supply), for cathodes
and anode wires. In particular,
automatic tripoffs of wire planes due to transitory overcurrents 
generated an alarm in the control room. 
\newline
Such trips occured on a very 
small number of planes for which the electrostatic insulation had some
weaknesses, very difficult to localize as the overcurrent appeared after
several hours or days of perfect operating conditions. Only 2 such planes
out of 147 had to be permanently switched off towards the end of 
the data taking.

The quality of the alignment (subsection~\ref{sec:reso}) 
for the chambers was checked daily with
a sample of 10,000 muons, and a complete realignment was done when
a wider residual distribution was observed.
Typically each
yearly run was divided in 5 to 10 periods with different alignment parameters.
This was found sufficient to adequately take care
of slow changes in the gas composition as well as of significant 
variations of the
atmospheric pressure during some periods.

%
%
\section{The drift chamber performances}
\label{sec:alignement}

  The chambers are operated at 100 V above the beginning of the
efficiency plateau, which was found to be 200 V wide. 
  Under these conditions, the typical wire
  efficiency is 0.97, most of the loss being due to the wire supporting rods.
The TDC's measure a time difference with a common stop technique:
$$\Delta t = t_{stop} - t_{start},$$
where
$$t_{start} = t_{interaction} + \Delta t_{tof} + \Delta t_{drift}
          + \Delta t_{wire} + \Delta t_{elect}$$
which takes into account the interaction time ($t_{interaction}$),
the time of flight
of the measured particle between the interaction point and
the measurement plane ($\Delta t_{tof}$), the drift time of electrons 
($\Delta t_{drift}$),
the signal propagation time\footnote{The speed of the signal propagation
along the wire was measured to be $\rm 26\ cm/ns$.}
 along the sense wire ($\Delta t_{wire}$) and
the signal propagation time in the electronic circuits ($\Delta t_{elect}$).
$$t_{stop} = t_{interaction} + \Delta t_{tof\_trig} + \Delta t_{delay}$$
consists of the time of flight of the triggering particle
($\Delta t_{tof\_trig}$)
and a delay ($\Delta t_{delay}$) long enough that the stop signal 
arrives after all possible start signals. $\Delta t_{delay}$ is of the 
order of one microsecond.
The time measured by a TDC is then
$$\Delta t = (\Delta t_{tof\_trig} + \Delta t_{delay})
           - (\Delta t_{tof} + \Delta t_{drift} +
              \Delta t_{wire} + \Delta t_{elect})$$
This time can be corrected for $\Delta t_{delay}$
 (see subsection~\ref{sec:tdc}),
 $\Delta t_{elect}$ and $\Delta t_{wire}$
(as soon as the $x$ coordinate of the hit along the wire is known), 
so that the true drift time is extracted.


The 
coordinate $c$ is obtained from the 
coordinate $c_w$
of the wire and from the drift distance $d$, which is a function of both
the drift time $\Delta t_{drift}$ and
the local angle $\phi$ of
the track in the plane perpendicular to the wires.
The drift distance is either added to or subtracted from
$c_w$. This is known as the up-down ambiguity:
$$c = c_w \pm d(\Delta t_{drift},\phi)$$
and this ambiguity is only resolved at the level of track reconstruction 
and fit.

\subsection{Spatial resolution} \label{sec:reso}
The spatial resolution 
depends on the precise knowledge of the 
wire positions and of the time-to-distance relation.
Since the wires 
are
glued at both ends and
on 3 vertical rods as mentioned in subsection~\ref{subsec:cell},   
the wire shape is approximated by 4 segments 
joining these
points. The vertical positions of 
these 5 points for each of the 6174 wires have then to be 
precisely determined.

One time-to-distance relation is defined for each 
plane independently, as different planes could work under different
gas\footnote{Due to different nitrogen contamination caused by different
leak rates.} and high voltage conditions. 
The
precise determination of 
the drift velocity\footnote{No dependence of drift velocity on the presence of
the magnetic field (after applying the compensating voltage on strips)
was found in our data.} 
is important because a mistake 
of one percent leads to a bias
in the spatial resolution of the order of several hundred microns
($d_{max} \cdot \frac{\Delta V_{drift}}{V_{drift}} = {3.2 \ \rm{cm}} \cdot 0.01 = 320 \ \mu$m).

A special alignment procedure~\cite{Kyan} 
was set up to measure wire positions and shapes, 
time-to-distance relationship and other relevant quantities
using muons crossing the NOMAD detector.

Track reconstruction details are given in the next section.
Our alignment procedure is based on the residual distributions of
reconstructed tracks (a residual is the difference between 
the hit coordinate and the reconstructed track coordinate at a
given measurement plane).
Initial wire positions are first taken from the geometers' survey and
the mechanical measurements, and then
corrected to minimize any systematic offset in an
iterative 
way:
typically 3 to 5 iterations on 100,000 muons samples are 
needed 
for the procedure
to converge.

\begin{figure}
 \begin{center}
 \hspace{0mm} \epsfig{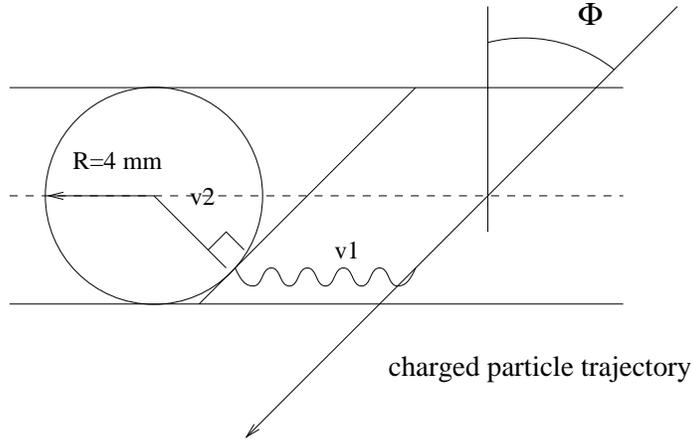}
 \end{center}
 \caption{A model used to describe the drift of electrons in the drift cell.}
 \label{fig:dc_t2d}
\end{figure}

The procedure then focuses on the time-to-distance relation.
A model used to describe the
main time-to-distance relation 
is
shown schematically in 
figure~\ref{fig:dc_t2d}: ionization electrons first drift
parallel to the strip plane with velocity $V_{1}$ and then radially 
to the wire with velocity $V_{2}$. For each plane these 2 velocities 
are
extracted again by minimizing offsets in signed residual distributions.
The dispersion of $V_{1}$ values measured for different planes 
was found to be 1.7\%.

The $z$ positions of the measurement planes are also 
updated
during this global
alignment procedure.

\begin{figure}
 \begin{center}
 \hspace{0mm} \epsfig{file=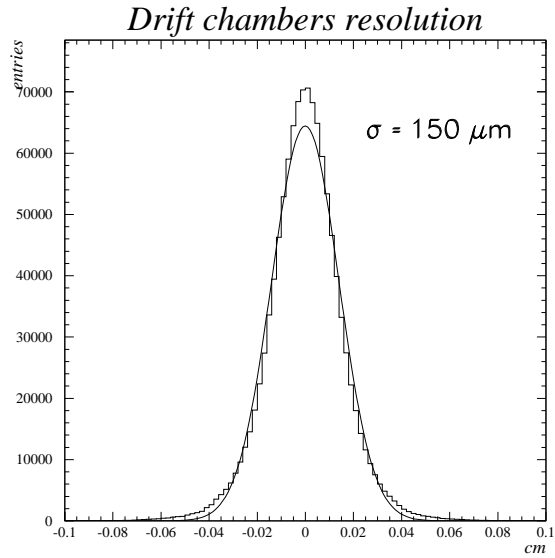,width=80mm}
 \end{center}
 \caption{Residuals for a sample of normal incidence tracks similar to the ones used for the alignment of the drift chambers.}
 \label{fig:dc_resid}
\end{figure}

\begin{figure}
 \begin{center}
 \hspace{0mm} \epsfig{file=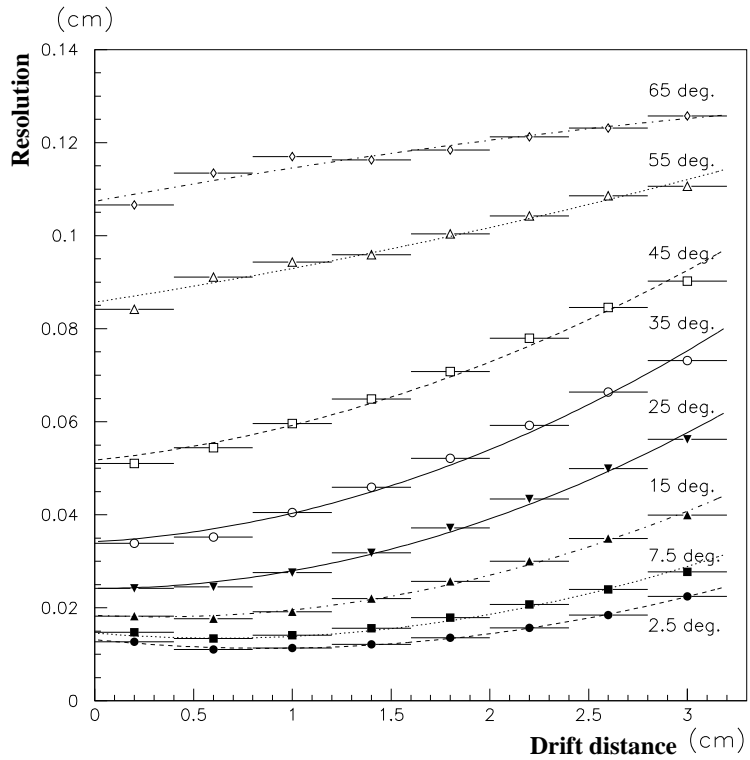,width=100mm}
 \end{center}
 \caption{The dependence of the track residuals on the drift distance 
for different
crossing angles.}
 \label{fig:dc_angles}
\end{figure}

At the end of this procedure (typically 10-15 iterations over samples of 100,000 muons)
the distribution of the residuals for tracks perpendicular to the chambers
is shown in figure~\ref{fig:dc_resid}.
  This distribution 
  has a $\sigma$ of about 150 $\mu$m (for normal incidence tracks) 
and confirms the good spatial resolution
  of the NOMAD drift chambers. The spatial resolution averages to 200 $\mu$m
for tracks originating from neutrino interactions (the average opening angle
is 7 degrees).
  The dependence of the resolution on the drift distance and the crossing
  angle $\phi$ is shown in figure~\ref{fig:dc_angles}.
At small drift distances, the angular dependence is essential due to the
electronic threshold (see subsection~\ref{sec:preamplifier}). 
At large drift distances, because of the non-uniformity 
of the electric field near the strips, the angular effect is enhanced.

\subsection{Efficiency}

The drift chamber efficiency and its dependence on the track angle and track 
position in the drift cell
were carefully studied using muons crossing the detector
between two neutrino spills.

\begin{figure}
 \begin{center}
 \hspace{0mm} 
 \epsfig{file=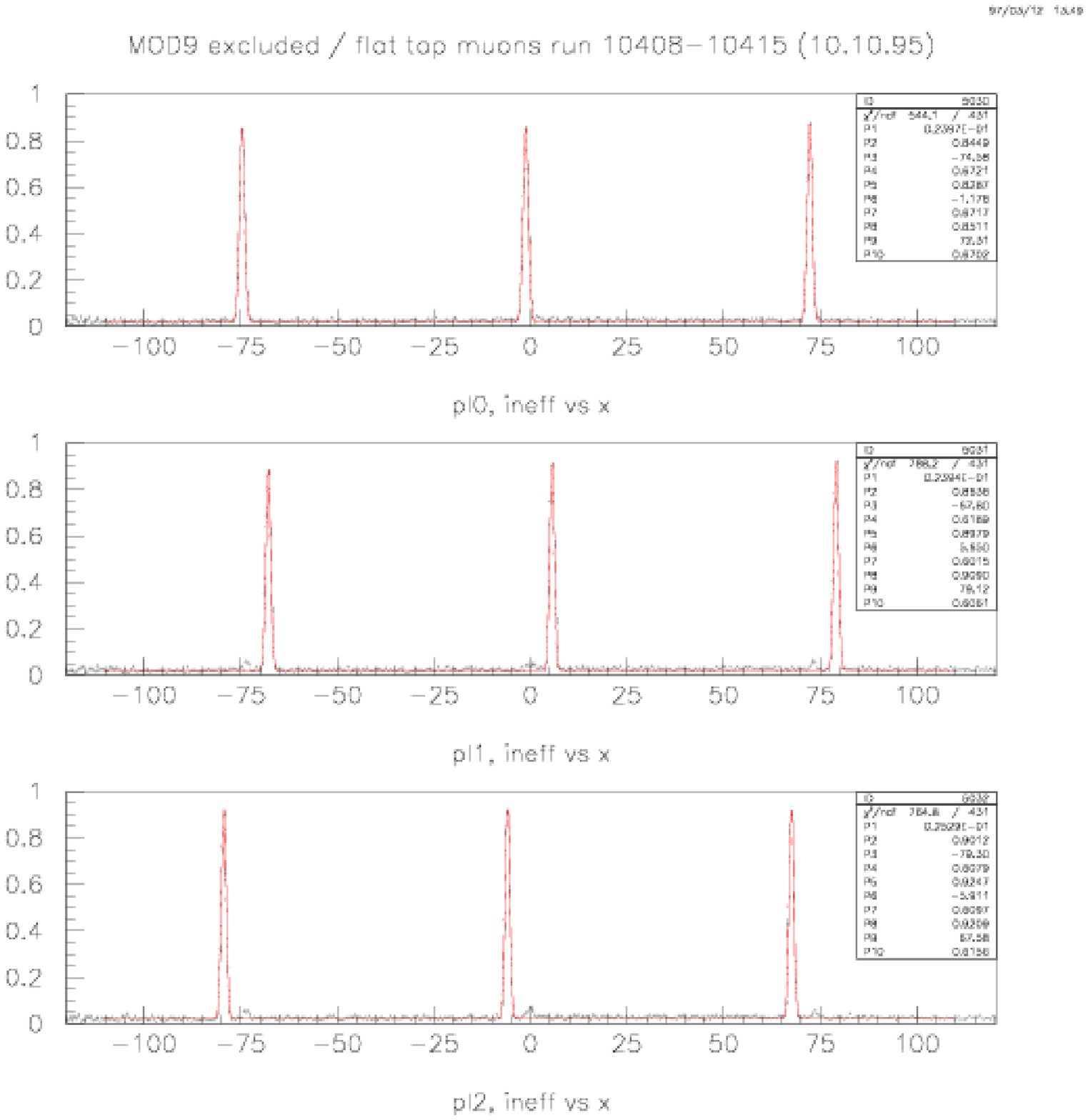,width=140mm,height=180mm}
 \end{center}
\vspace{-2cm}
\caption{
The inefficiency (1-$\varepsilon_{min}$) as a function of 
the $x$-coordinate (along the wire). 
Peaks correspond to the wire supporting rods.
The small bumps in the 
second (pl1) and third (pl2) planes are due to spacers used as chamber
supports.
}
\label{fig:dceff_bar95_norm}
\end{figure}

The inefficiency  was computed as a function of 
the $x$-coordinate (along the wire). The results are given in 
figure~\ref{fig:dceff_bar95_norm}. This distribution can be well fitted
by a constant ($\approx 2.4\%$) and three gaussian functions with a width of
$\approx 6$ mm centered at the supporting rod positions.  
As a result, 
the efficiency in the region between 
supporting rods is 
$\varepsilon_{min} \approx 97.6\%$ 
consistent with our expectations
and we confirm that 
the inefficiency is caused mainly by the presence of the rods.

\begin{figure}
 \begin{center}
 \hspace{0mm} 
 \epsfig{file=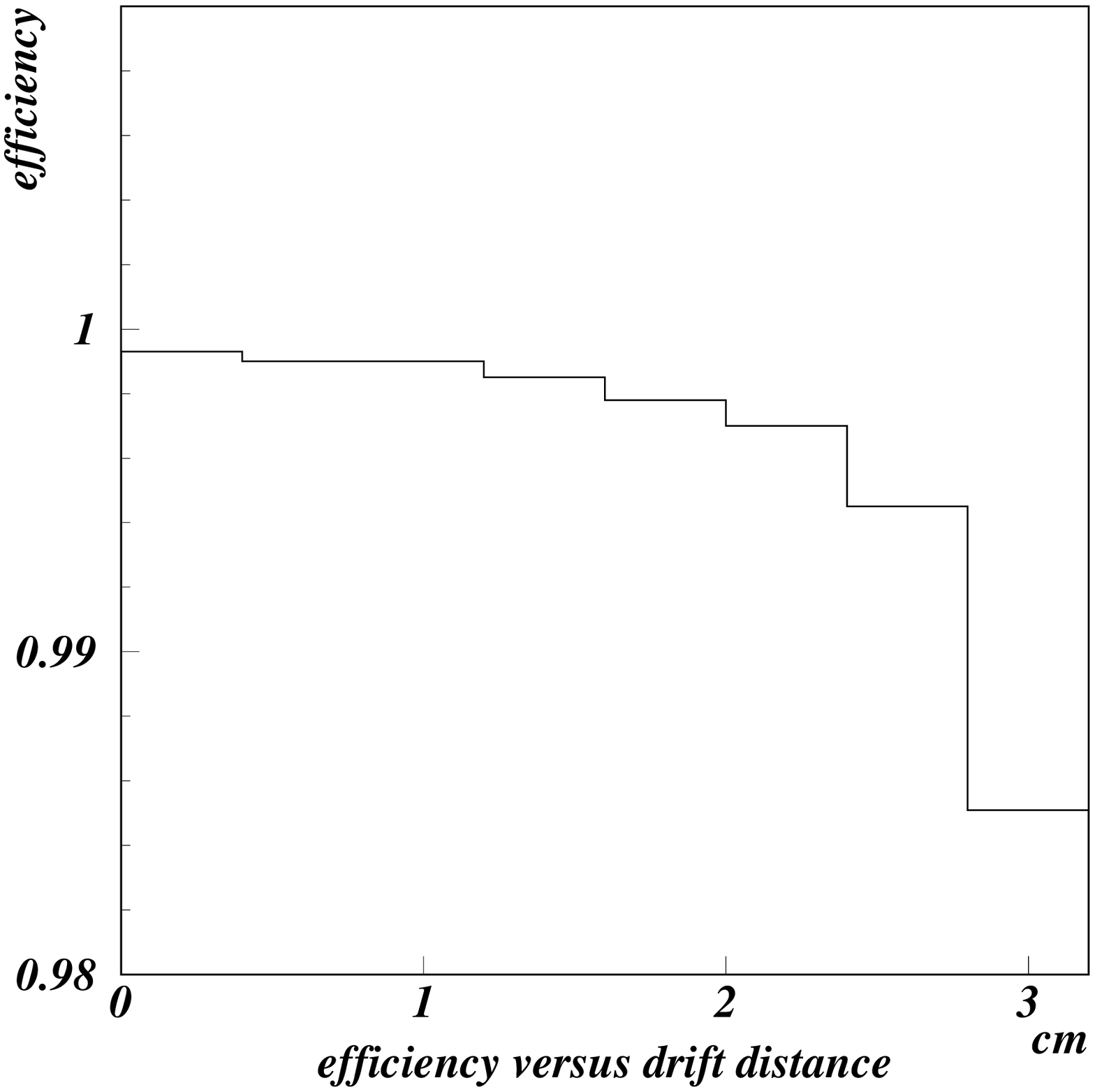,width=60mm,height=50mm}
 \end{center}
\caption{
The dependence of the hit finding efficiency on the track position 
in the drift cell.
}
\label{fig:dceff_caren_ineff_regions}
\end{figure}

Further studies show that the major efficiency loss is not due to the
absence of the hit in a measurement plane but due to the non-Gaussian tails
in the residual distributions. If one extends the road for the hit 
collection during the track reconstruction a maximal drift chamber
efficiency of $\varepsilon_{max} \approx 99.7\%$ can be obtained. 
One can also study the efficiency as a function of the track position in the drift cell
(figure~\ref{fig:dceff_caren_ineff_regions}): the main loss occurs at the edge
of the cell where the drift field is less uniform (see  figure~\ref{fig:dc_BonBoff}) .

\begin{figure}
 \begin{center}
 \hspace{0mm} 
 \epsfig{file=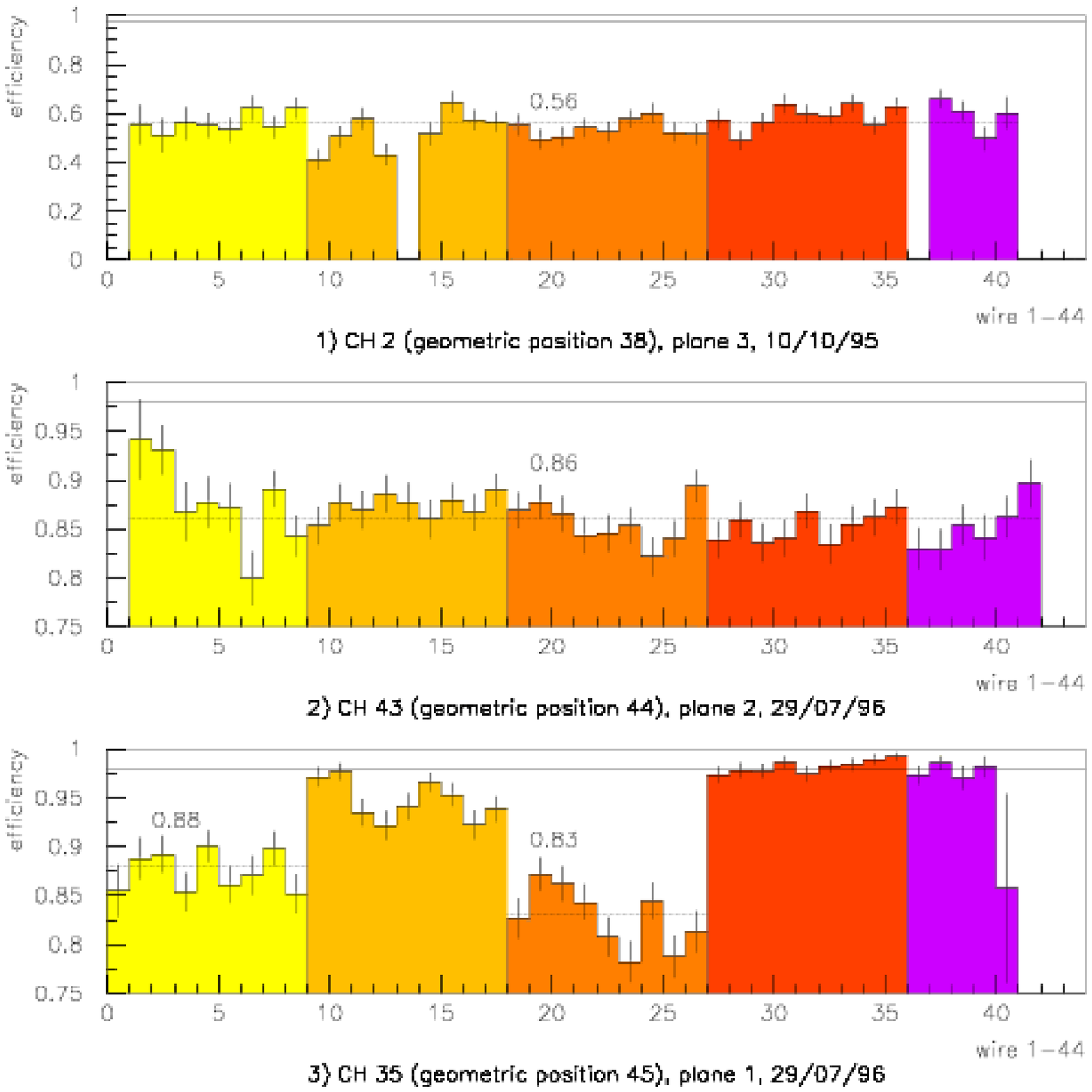,width=100mm,height=120mm}
 \end{center}
\caption{
Examples of hardware efficiency losses for some pathological
drift chamber planes.
The horizontal solid line shows an efficiency $\varepsilon = 0.98$ 
for normal chambers.
The different shaded areas indicate different 
strip bands.
(1) for planes with short circuits between strips: the efficiency loss is $\approx 42\%$;
(2) for disconnected field wires: the efficiency loss is $\approx 12\%$;
(3) for misaligned strip bands: the efficiency loss is $\approx 10 \div 15\%$.
}
 \label{fig:dceff_carenineff_hard}
\end{figure}

There are other hardware effects which 
cause efficiency
losses: planes with short circuits between strips, disconnected field wires,
misalignment of a wire with respect to the facing strip band.
These effects have been studied in details
(see figure~\ref{fig:dceff_carenineff_hard} for a particular example)
and the following typical values have been obtained:\\ 
- planes with short circuits between strips: efficiency loss $\approx 40\%$,\\
- disconnected field wires: efficiency loss $\approx 10\%$,\\
- misaligned strip bands: efficiency loss $\approx 10 \div 15\%$.

We did not notice any correlation of the loss in efficiency with oxygen 
contamination in the gas mixture or chamber leak rate.
 
As an example, the overall
hardware performance of the NOMAD drift chambers
during the 1996 run data taking period
was the following:
among 147 planes,
1 to 2 planes were switched off due to unrecoverable problems 
(such as a broken wire), 
2 to 3 planes suffered from short circuits between strips 
and 3 planes had disconnected field wires.


\subsection{Afterpulses}
\label{subsec:afterpulses}

\begin{figure}
\begin{center}
\mbox{\epsfig{file=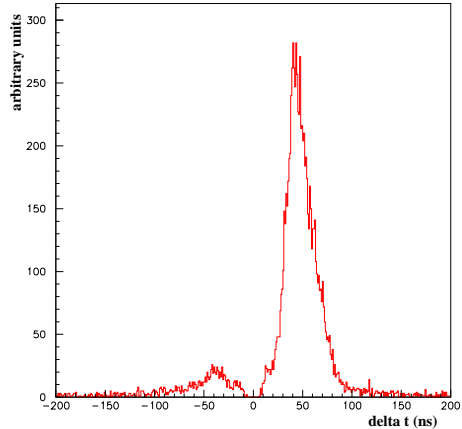,width=6cm,height=6cm}}
\end{center}
\caption{
Example of the time difference (in ns) between 
a second
hit on the same wire and 
the hit included in a given track. 
}
\label{fig:dt_examp}
\end{figure}

The problem of afterpulses (or bounces) was present in the drift chambers
response: 
some of the hits can
be accompanied by one or several other hits on the same wire. 
The time difference
between the hit included in a given track and the other hit on the same wire
is shown in
the 
distribution of~figure~\ref{fig:dt_examp}.
We have noticed
two contributions: the first one is symmetric in time and the second one
is concentrated at about 50 ns after the first 
digitization. 
The former contribution was attributed to the emission of low-energy 
$\delta$-rays which produce random hits in the drift cell crossed
by the track, while the latter contribution was associated with the
smearing of the electron cloud consisting of several clusters which 
could trigger another digitization after the arrival of the first electron.
It was found 
that the rate of 
afterpulses depends on the track angle and track position in the drift cell.
These afterpulses 
were included
in the simulation program.

The knowledge gained by the studies of resolution, efficiency
and the presence of afterpulses was used during the track reconstruction in the
drift chambers. The dependence of the resolution on the drift distance and
track angle was parametrized and implemented at the level of track search 
and fit.
A special bounce filter was developed to cope with the presence of afterpulses
for the hit selection.

%
%
\section{The drift chamber reconstruction software}
\label{sec:reconstruction}

In the NOMAD experiment trajectories of charged particles are reconstructed 
from the coordinate measurements provided by the drift chamber (DC) system.
The main purpose of the drift chamber reconstruction program is 
to determine the event topology and to measure 
the momenta of charged particles. 
The reconstruction program for the NOMAD drift chambers 
is extremely important for the performance and sensitivity of 
the experiment. A very high efficiency
of the track reconstruction is required in order to provide good measurement
of event kinematics for the oscillation search. We have also to be sure
that the measured track parameters do not deviate significantly from the true
particle momenta, 
i.e.
the reconstruction program should provide good
momentum resolution. 
The amount of ghost tracks should be minimized.
Since
in the NOMAD setup the amount
of matter crossed by a particle between two measurement planes cannot be
neglected, the effects of energy losses and multiple scattering must be
carefully taken into account. 

The task of the reconstruction program is two-fold. First, it should
perform pattern recognition (track search), namely to decide which individual 
measurements provided by the detector should be associated together to form 
an object representing a particle trajectory. 
At the next stage, a fitting procedure should be applied to this set of 
measurements in order to extract the parameters describing the 
trajectory out of which the physical quantities can be computed. 


The track finding procedure consists of two loosely coupled tasks: the first one guesses 
possible tracks from hit combinatorics, and provides 
initial
track parameters.
The second task attempts to build a track from the given parameters by repeatedly collecting hits, fitting and rejecting possible outliers. The track 
is claimed
to be fitted when no more hits can be added to it. We developed several 
approaches to the first task which 
are summarized in~\cite{Manu}.
A short overview is given here.

\subsection{Searching for candidate tracks}
\subsubsection{The DC standalone 
pattern recognition
}
The algorithm presented here does not make use of any prior information for 
the track search: 
the parameter space has 5 dimensions (or even 6 if one adds the trigger 
time jitter). The combinatorics from hits to likely track parameters goes 
in two steps:
we first associate hits from the 3 planes of the same chamber in triplets, 
which provide some 3-dimensional (3D)
information; we then search combinations of 3 triplets 
belonging to
the same helix.

{\it DC Triplets}\\
A drift chamber is made of 3 sensitive planes measuring the U, Y and V 
coordinates, where:
\begin{center}
$u \;= \; y \cdot \cos 5^\circ\;-\;x \cdot \sin 5^\circ$ 
and $v \;= \; y\cdot \cos 5^\circ\;+\;x \cdot \sin 5^\circ$
\end{center}

Four parameters define locally the track (across one chamber, 
one can neglect its curvature): we use 2 coordinates in the central Y plane  $x,y$
and 2 slopes $\alpha_x$, $\alpha_y$ (with respect to Z). 
From 3 hits in the 3 planes and 3 chosen signs
(we decide if the track crossed above or below each involved sense wire), we have
3 measured coordinates U, Y and V. With 3 measurements, we cannot estimate 
the 4 local parameters. The measured combinations are:
\begin{eqnarray*}
y & = & Y \\
2 \Delta z\ \alpha_x \sin 5^\circ & = & U\;+\;V\;-\;2 Y \cos 5^\circ \\
x + \Delta z\ \ cotg 5^\circ\ \alpha_y & = & (V-U)/(2 \sin 5^\circ)
\end{eqnarray*}

where $\Delta z$ is the distance between two consecutive measurement planes 
(about 2.5 cm). The second equation can be turned into a constraint
by assuming that $\alpha_x$ does not exceed a certain bound (e.g. $| \alpha_x | < 1$). Since $\sin 5^\circ \simeq 0.087$, this constraint becomes the first tool
to assemble triplets. The third equation shows that $x$ remains poorly known as long as the $y$-slope ($\alpha_y$) of the track is unknown: changing $\alpha_y$ by 1
shifts $x$ by about 30 cm.

To build triplets, we compute $\Delta = U\;+\;V\;-\;2 Y \cos 5^\circ $ for all hits
and sign combinations inside a chamber, and solve 
ambiguities by keeping
no more than 2 triplets per signed hit, possible triplets being
ranked by increasing $|\Delta|$.

{\it Helix search}\\
Triplets are used to initiate the track search. With 3 triplets 
we are
able to test whether they belong to the same trajectory.
At this step,
we still ignore the magnetic field variations
and check the 3 triplet combinations against a perfect helix.
From the 3 $y$ positions, we compute 3 slopes in the YZ plane
which enable to compute 3 $x$ coordinates from the third equation above.
We 
can
then 
reliably compute drift 
distances, including a sensible time correction due to the signal propagation
along the sense wire 
and
using the known track slope for the time-to-distance
relation
(see subsection~\ref{sec:reso}).
We then
recalculate
the 3 triplet ($x,y$) positions from these precise drift 
distances
and slopes.

The most sensitive criterion is to check by how much the $x$ coordinates depart
from the same helix. Calling $\phi$ the angle with respect to (e.g) 
horizontal in the YZ plane, the residual is:
\begin{center}
$x^{extrap} -x_2 = x_1 + (\phi_1 - \phi_2) \cdot \frac  {x_3 - x_1}{\phi_3 -\phi_1}-x_2$
\end{center}
where the subscript refers to the triplet number. 
For accepted combinations, we compute the track parameters at the central 
triplet position
and feed it to the track construction task.

\subsubsection{The coupled TRD and DC track search}
The TRD detector~\cite{NOMAD_TRD} is located downstream of the target and 
has a total length of 1.5 m in Z.
Five drift chambers are interleaved in the TRD modules, which measure the 
$x$ coordinate
of tracks by means of vertical cylindrical xenon filled straws, 1.6 cm in 
diameter without drift time information. 
These additional drift chambers allow
to improve the
momentum resolution for reconstructed tracks and provide more precise
track extrapolation to the preshower and electromagnetic calorimeter
front face. 

The first step 
of the coupled TRD-DC track search algorithm~\cite{Manu}
consists in reconstructing TRD
tracks (in XZ projection) using the 9
planes of straws. The coarse spatial sampling enables to ignore the effects
of curvature in the XZ plane: a TRD track defines a position and a slope in the XZ plane. These TRD tracks can only be used in the most downstream part of the target,
and this TRD seeded track search only applies to the 10 
most downstream drift chambers. In those
chambers, and for every TRD track, we can build
triplets using the $x$ and $\alpha_x$ values,
thus
triplets are now 
constrained. We can even 
build doublets, a combination of two hits in a given chamber. 
The helix search can
then be carried out in the same way as for the DC standalone case. 
In a later version,
we slightly improved the reconstruction quality of complicated events
 by searching circles among DC hits projected on a vertical plane 
containing the TRD track. This was made possible by the increase of CPU power
of low cost computers.

\subsubsection{Track search using vertex information.}

Advanced track construction algorithms which take into account the information
about reconstructed vertices\footnote{The vertex reconstruction 
is described in subsection~\ref{sec:vertex}.}
have also been developed~\cite{Nathalie}. 

{\it Tracks from 1 vertex}\\
Searching a track that emerges from a vertex reduces the parameter space dimension by 2 units.
Since the vertex is already a 3D point (much better defined than any triplet), 
we can
find
a short track which only has 2 triplets.
The triplets are here searched for with $x$, 
$\alpha_x$, $y$ and $\alpha_y$ 
loosely constrained to point back to the vertex at hand. This enables to
find tracks at large angles with respect to the beam direction. 
Having constructed these triplets,
the triplet combinatorics for helix search can be run in the same way as 
described previously.
The vertex is considered as a 
triplet 
for which the
$x$ position does not depend on the YZ slope. 
The 
new
track is only accepted if it enters the seed vertex with an acceptable
$\chi^2$ increase.

{\it Tracks joining 2 vertices}\\
Hadron interactions may be reconstructed as secondary vertices or hanging
secondary tracks without the primary hadron being found. We designed a 
track search algorithm
to find tracks either between 2 actual vertices (i.e. vertices with at least
2 attached tracks) or between an actual vertex and 
a starting point of a standalone track
(in which case
its actual vertex may slide along the track direction). This is a 
track search with 1 or 2 free track parameters which does not go through triplets.

\subsection{Building tracks}

This task uses track parameters (given by the previous track search) at 
any given
plane as a seed and 
tries to build a track.

The first step consists in collecting hits upwards and backwards from a given plane
within a road (typically 3 mm). The hit search makes use of all available
information by computing drift distances corrected for slope and $x$ position. The
track global time offset with respect to the trigger time is still unknown and 
its average value
is taken first.
Later on,
the time offset of the highest momentum track already found 
is used for all the tracks in the event.
The collection stops when too many measurement planes 
are crossed without a matching hit or when several hits
are found within a control road in the same plane. 
Missing hits can be ``excused'' if
the wire or plane at the expected position is dead, or if a hit with a smaller
drift time on the wire may hide the expected digitization. A track
may go to the next step if it has enough hits, and a high enough average efficiency.
At this last step, we first fit the hit list. The candidate track disappears
if the fit diverges or if the $\chi^2$ is too high. We then 
discard
hits
that exhibit a too high $\chi^2$ contribution (the cut is usually 10).
From now on, the track gets re-fitted whenever we add one hit, and 
the $\chi^2$ increment decides whether a hit may enter a track.
Using the Kalman filter technique makes this approach acceptable in
terms of CPU. We then try to collect hits in planes 
crossed by the track but where hits are originally missing. We finally
collect hits downstream and upstream, and iterate collection
over the track,
downstream and upstream until the track hit list stops evolving.
We store the track in the 
track repository, and mark its hits
as no longer available for triplet construction or hit collection by other tracks.
They however still remain examined for the hit ``excuse'' mechanism.

\subsection{Track model}

The track model describes first the dependence of the measurements on the initial
values in the ideal case of no measurement errors and 
of deterministic interactions of a particle with matter.
During its flight through the detector a particle however encounters various
influences coming from the materials of which the detector is built.
There are effects which can be taken into account in a deterministic
way: average energy loss and average multiple scattering. They depend in general 
on the mass of the particle, its momentum, the thickness and nature of the 
traversed material.
The detailed information on the track model used for the track reconstruction
in the NOMAD drift chamber system can be found in~\cite{Boris}.

The track model has been used to develop an extrapolator package which is
heavily used for the track construction and fit~\cite{Boris}: 
there, a precise track model and a good magnetic field description are 
important 
to ascertain that
the fitting procedure
is unbiased. 
The magnetic field inside the magnet has been carefully measured and 
parameterized. In the detector fiducial volume the main component of 
the field $B_x$ has been found to vary within 3\%. 
Uncertainties in the track model predictions due to small non-zero values of 
the two other components ($B_y$ and $B_z$) are negligible compared to 
the effect of multiple scattering. 
This allows to use only the local value of the main field
component $B_x$ for the track model calculations~\cite{breakpoint}.


\begin{figure}
\begin{center}
\epsfig{file=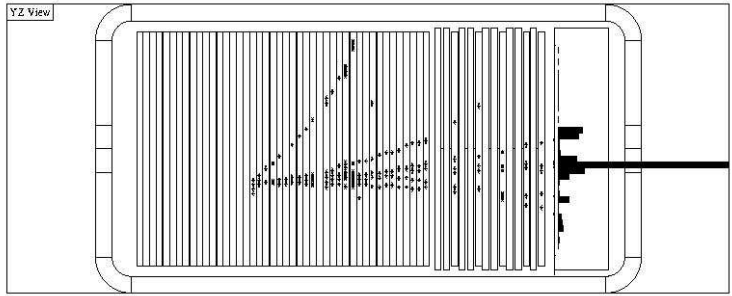,angle=-90,width=150mm}
\caption{
Raw hits before the reconstruction in a real data event.
One can easily see a dead
drift chamber plane in the middle of the detector. 
}
\label{raw_data}
\end{center}
\end{figure}

\begin{figure}
\begin{center}
\epsfig{file=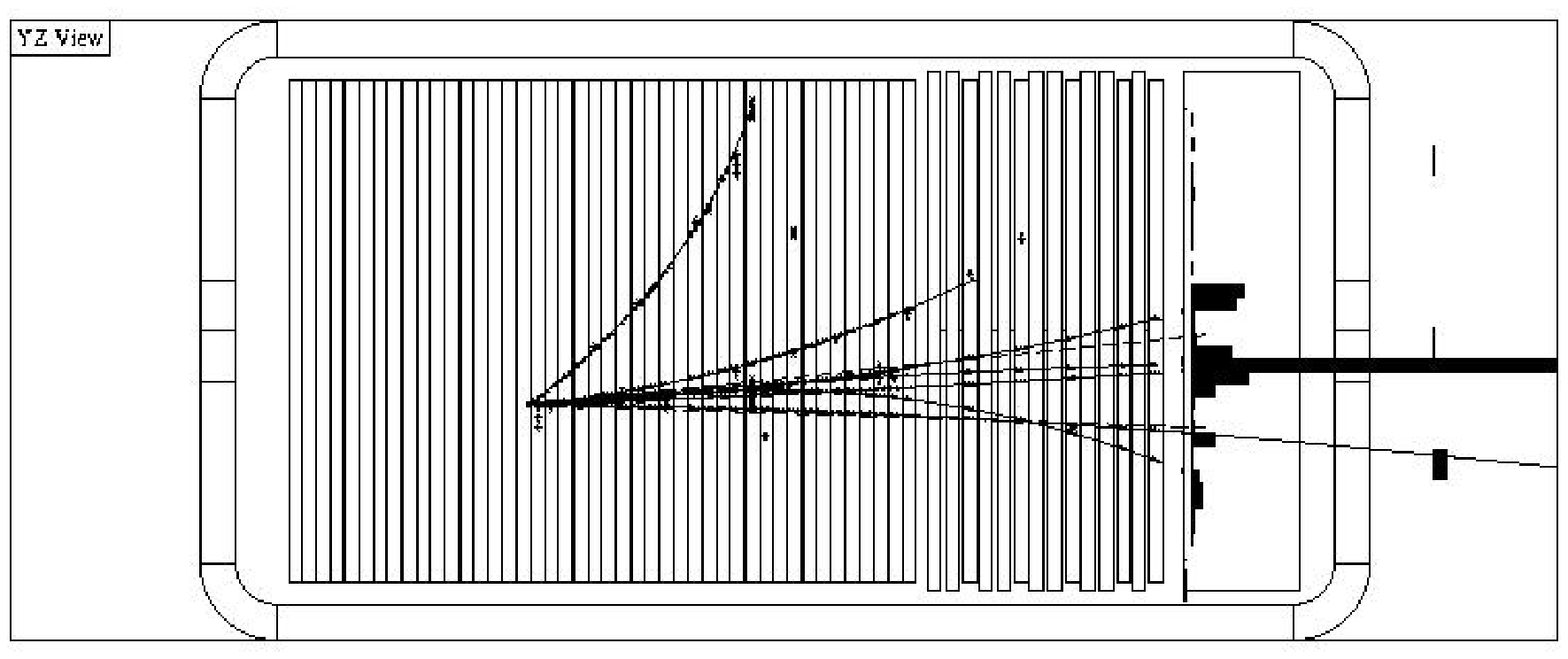,angle=-90,width=150mm}
\caption{
The same event as on figure~\ref{raw_data} 
after performing DC reconstruction
and particle identification.
}
\label{rec_data}
\end{center}
\end{figure}

\subsection{Track fit}

After pattern recognition, one has to determine the track parameters, 
in particular
charged particle momenta. 
The Kalman filter technique~\cite{fruhwirth,BilloirEtAl} 
is adapted to fulfill this task.
Some details on the implementation of the Kalman filter 
for the NOMAD drift chambers can be found in~\cite{breakpoint}.

The track fit proceeds in 3 steps: forward filtering, backward filtering and
smoothing~\cite{fruhwirth}. 
The smoothing provides the best possible track position estimate at
any measurement location, thus allowing 
to efficiently remove wrong associations.

The fitting routine performs those 3 steps 
until the $\chi^2$ modification between 2 fits is below a typical value of 0.1.
 
The Kalman filter technique was implemented in two different ways: using weight
and covariance matrices. The latter was found faster since it allows to 
avoid several matrix inversions per hit when the effect of 
multiple scattering is taken into account.

A raw event from real data can be seen in the
figure~\ref{raw_data}.
The result of DC reconstruction and particle identification 
is shown in figure~\ref{rec_data}.

An algorithm developed on the basis of the Kalman filter technique 
to search for
potential break points corresponding to hard bremsstrahlung photons emission is
discussed in~\cite{breakpoint}.

\subsection{Vertex reconstruction} \label{sec:vertex}

\subsubsection{Vertex finding and fitting}

The vertex reconstruction 
is performed with reconstructed tracks. 
The major tasks of the vertex package are:
\begin{itemize}
\item to determine the event topology (deciding upon which tracks should belong
to which vertex);
\item to perform a fit in order to determine the position of the vertex and 
the parameters of each track at the vertex;
\item to recognize the type of a given vertex (primary, secondary, $V^0$, etc.).
\end{itemize}

The vertex search algorithm proceeds as follows. For both 
ends
of each track one calculates the minimal weighted distances 
to
all the other
tracks.
Then one 
takes the point corresponding to the
minimal 
distance as a starting vertex position; tracks close enough to the chosen point
are added to the list of 
tracks of the candidate vertex.
A fit algorithm (Kalman filter) is applied next to reject unmatching
tracks. Finally, one defines the topology of the reconstructed vertex
assuming the direction of the incoming 
particle
and the balance of momenta
between tracks belonging to this vertex: 
e.g., a track connected by its end to a vertex
could be either a parent track (scattering vertex or decay of a charged
particle) or a track going backwards (in that case it has to be reversed,
i.e. refitted taking into account energy losses in the right way).
Additional vertices are searched for and reconstructed 
starting from
unused 
track 
ends.
The Kalman filter technique is used to allow fast vertex fit and provides 
a simple way 
to add or remove tracks from an existing vertex without completely 
refitting it.

\subsubsection{Vertex position resolution}
 
\begin{figure}
\begin{center}
\hspace{0mm} \epsfig{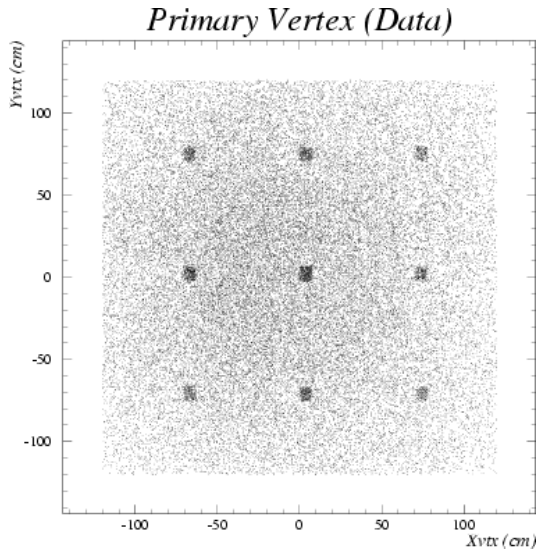}
\end{center}
\caption{The positions of reconstructed neutrino interaction vertices
		  from experimental data
          in a plane perpendicular to the beam direction; see text for details.}
\label{fig:dc_vertex_xy}
\end{figure}

Most neutrino interactions in the active target occur in the passive
panels of the drift chambers. 
Figure~\ref{fig:dc_vertex_xy} shows the distribution
of primary vertices in a plane perpendicular to the beam. A fiducial cut of
$ -120 \leq ~{x,y} ~\leq 120$~cm is imposed. The gradual decrease of 
the beam intensity with radius can be easily seen. The 9 dark
spots of high intensity are caused by the spacers which are inserted in the 
chambers in order to increase their rigidity and maintain the gap width
(see subsection~\ref{sec:panels}).

\begin{figure}
\begin{center}
\hspace{0mm} \epsfig{file=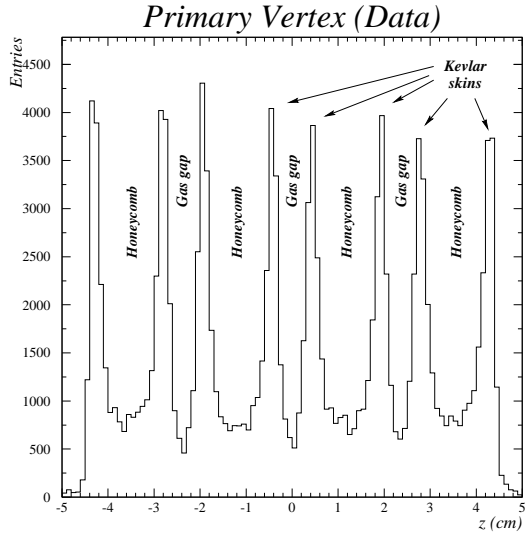,width=70mm,angle=-90}
\end{center}
\caption{The positions of reconstructed neutrino interaction vertices
		  in the experimental data
          as measured along the beam direction; see text for details.}
\label{fig:dc_vertex_z}
\end{figure}

Figure~\ref{fig:dc_vertex_z} shows the distribution of primary vertices along 
the beam direction. The information from 
all
drift chambers 
has 
been folded to cover the region of $\sim$10~cm around 
the centre of each chamber.
One can easily see that the bulk of neutrino interactions occurs in the walls
of the drift chambers.
The eight spikes in this distribution correspond to the
kevlar skins of the drift chambers (see figure~\ref{fig:dc_cut}). 
Regions in $z$ with a smaller number of reconstructed vertices 
correspond to the honeycomb panels and the three gas-filled 
drift gaps.

\begin{figure}
\begin{center}
\hspace{0mm} \epsfig{file=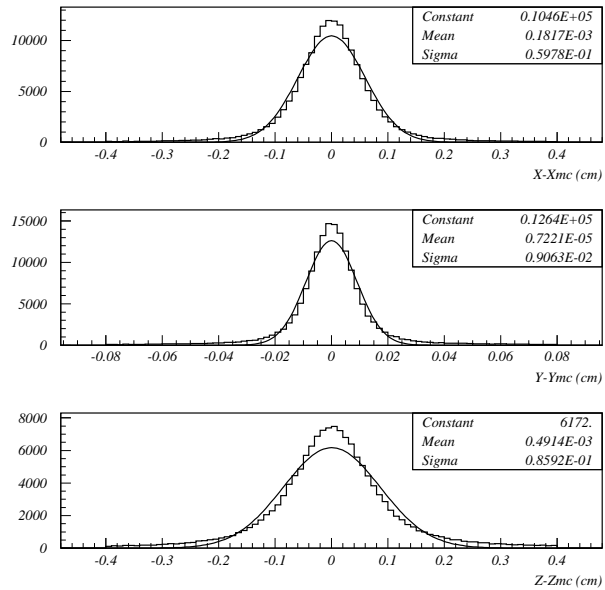,width=90mm}
\end{center}
\caption{
 The vertex position resolution for reconstructed $\nu_\mu$ CC MC events.
 The resolution is 600 $\mu$m, 90 $\mu$m and 860 $\mu$m for $x$, $y$ 
 and $z$ respectively.
}
\label{fig:dc_vertex_reso_mc}
\end{figure}

The vertex position resolution was checked using MC simulation. The results are
presented in figure~\ref{fig:dc_vertex_reso_mc}. 
Resolutions of 600 $\mu$m, 90 $\mu$m and 860 $\mu$m 
in $x$, $y$ and $z$ respectively are achieved.

\subsection{Implementation}
The drift chamber reconstruction is a software package written in C language 
in an object oriented way. It includes track and vertex search and fit, 
track extrapolation package and a graphical display which was found very 
useful to check the performances of the pattern recognition algorithms and 
to study possible improvements.

\subsection{CPU considerations}
The overall CPU time needed for the event reconstruction in the drift chambers
strongly depends on the complexity of the event. Events with
more than 1000 hits are not reconstructed. Genuine neutrino interactions
are reconstructed in  about 10 s on a PC at 300 MHz. Half of the time is spent
in combinatorics, the other half in the extrapolation of candidate track parameters
(and of the error matrix when needed) which is necessary when collecting
hits and fitting tracks.

The reconstruction algorithms have been explicitely optimized to reduce 
the CPU time required. However, this optimization became less critical
with a very fast increasing performance of low cost computers 
along the duration of the experiment.

%
%
\section{Check of the drift chamber performances using experimental data}
\label{sec:physique}

It was important to validate the claimed performances of the drift chambers
using experimental data.

\subsection{Momentum resolution}

\begin{figure}
 \begin{center}
 \hspace{0mm} 
 \epsfig{file=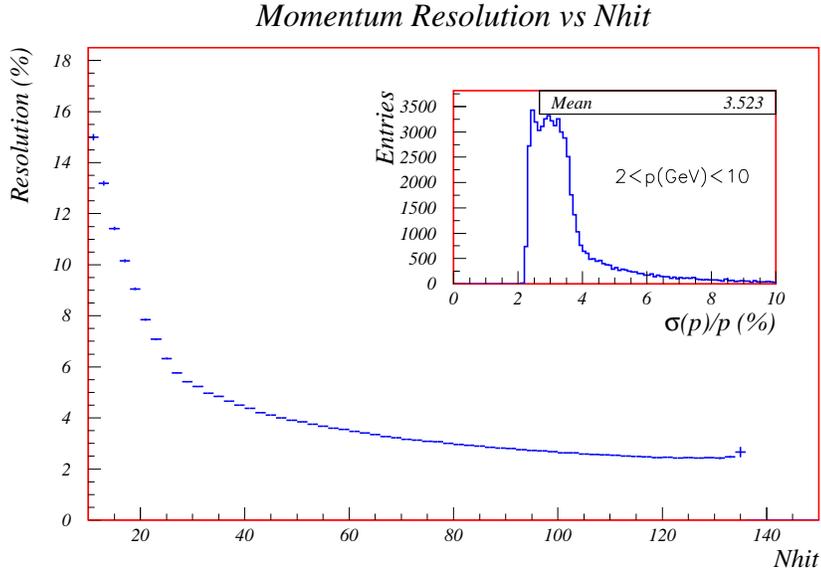,width=120mm}
 \end{center}
 \caption{
Average momentum resolution as a function of the track length
(number of hits in the fiducial volume of drift chambers).
The insert shows the distribution of the number of tracks at a given resolution.
}
 \label{fig:dc_reso_real}
\end{figure}

  The momentum resolution provided by the drift chambers is a function of
  momentum and track length. For charged hadrons 
  and muons travelling normal to the plane of the chambers, it
  can be approximated by 
$$
  \frac{\sigma_p}{p} \approx \frac{0.05}{\sqrt{L}} \oplus 
  \frac{0.008 \times p}{\sqrt{L^5}} 
$$
  where the momentum $p$ is in GeV/c and the track length $L$ in m.
  The first term is the contribution from multiple scattering and the
  second term comes from the single hit resolution of the chambers.
  For a momentum of 10 GeV/c, the multiple scattering contribution
  is 
the larger one 
as soon as
the track length is longer than 1.3~m. 

 Figure~\ref{fig:dc_reso_real} shows the resolution as a function of 
 the number of hits (related to the track length)
 as obtained from a fit of real tracks. A momentum
 resolution of $3.5\%$ in the momentum range of interest 
($\rm 2 < p (GeV/c)  < 10 $) is achieved.

  For electrons, the tracking is more difficult because they radiate
  photons via
  the bremsstrahlung process when 
crossing
the tracking system.
  This results sometimes in a drastically changing curvature.
  In this case, the momentum resolution as measured in the drift chambers 
is worse and 
  the electron energy is measured by combining information from the drift 
  chambers and the electromagnetic calorimeter~\cite{NOMAD_ECAL,Boris}. 

\subsection{Neutral strange particles} 

A study of neutral strange particles can provide some information related 
to the performance of the drift chambers.

\begin{figure}[htbp]
\begin{center}
\epsfig{figure=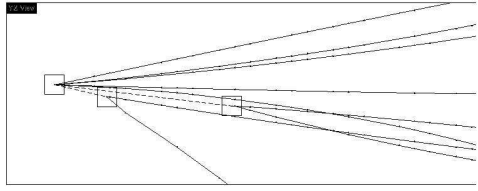,angle=-90, width=.7\linewidth}
\end{center}
\caption{
A reconstructed data event (zoom on the primary vertex) containing
2 \vo~vertices identified as \lam~and \alam~decays.
The scale on this plot is given by the size of the vertex boxes
($3\times3$ cm$^2$).
}
\label{Event_lam_alam}
\end{figure}

A decay of a neutral strange particle appears in the detector as 
a \vo-like vertex: two tracks of opposite charge
emerging from a common vertex separated from the primary neutrino interaction
vertex 
\cite{Kyan,Patrick,Cyril}. 
Figure \ref{Event_lam_alam} shows an example of a reconstructed data 
event with two such \vo's.

\begin{figure}[htbp]
\centerline{\epsfig{figure=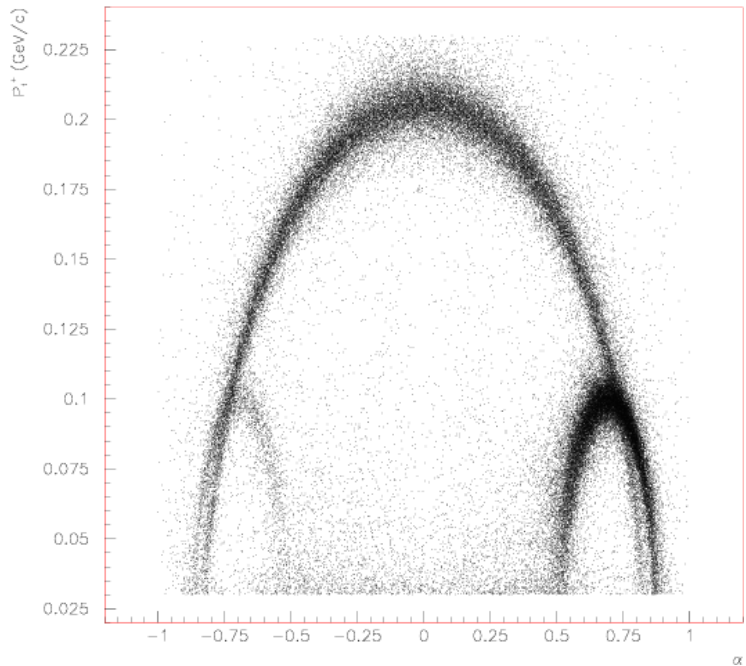, width=.7\linewidth}}
\caption{
Armenteros' plot for reconstructed \vo's: \ko's cluster on the large central ellipse, \lam's on the small right ellipse and the smaller sample of \alam's 
on the left one.}
\label{armenteros}
\end{figure}

The quality of the reconstruction in the NOMAD drift chambers allows 
a precise determination of the \vo ~decay kinematics as can be seen on 
the so-called Armenteros' plot (figure \ref{armenteros}). This 
figure is obtained by plotting for each neutral decay
the internal transverse momentum 
($P_t^+$) versus \al, 
the asymmetry of the longitudinal momenta of the two outgoing tracks
($\alpha = \frac{P_l^+ - P_l^-}{P_l^+ + P_l^-}$). 
Without any cut, each type of neutral strange particle appears clearly on the figure.

\begin{figure}[htbp]
\centerline{\epsfig{figure=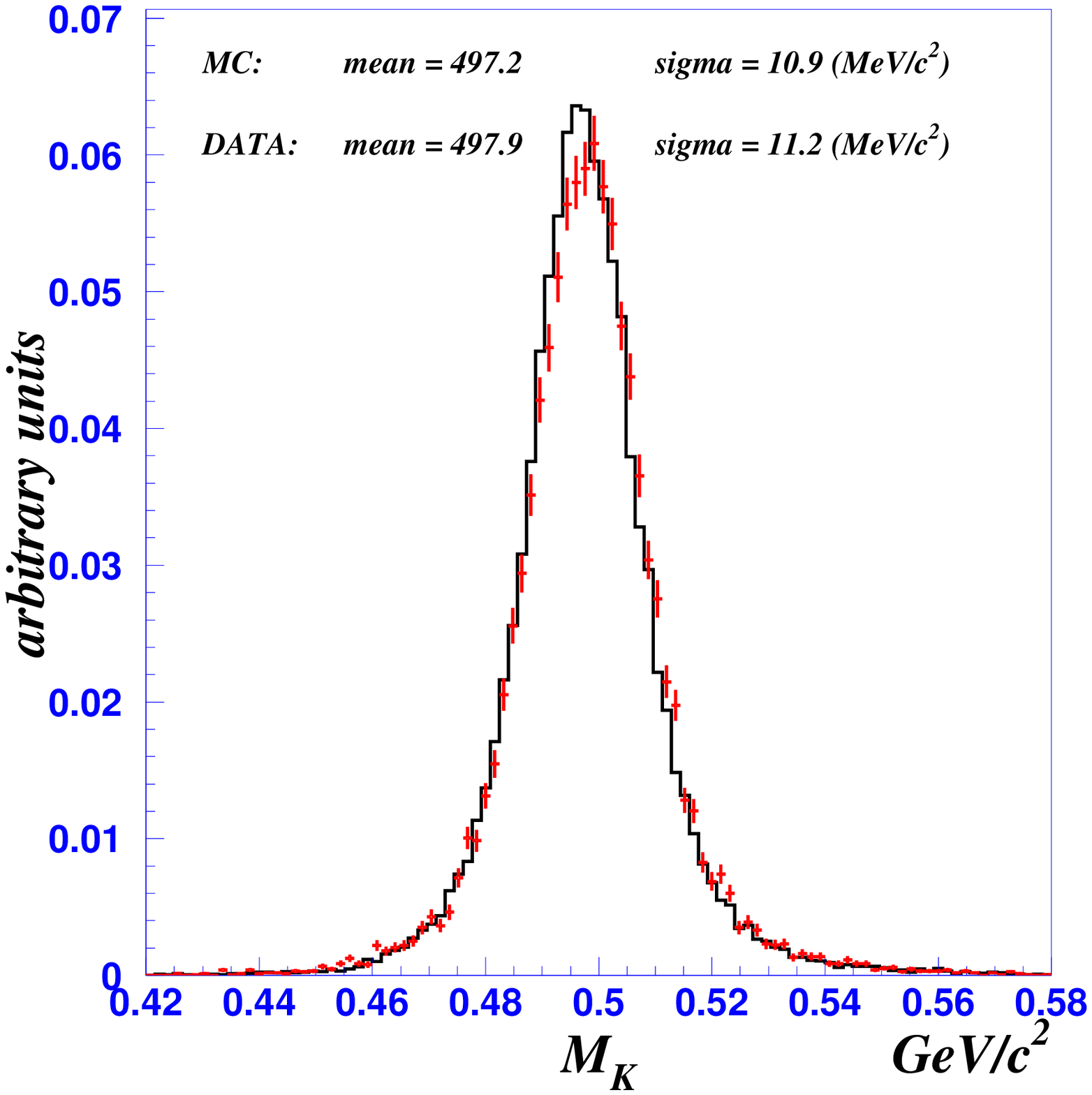, width=.49\linewidth}\epsfig{file=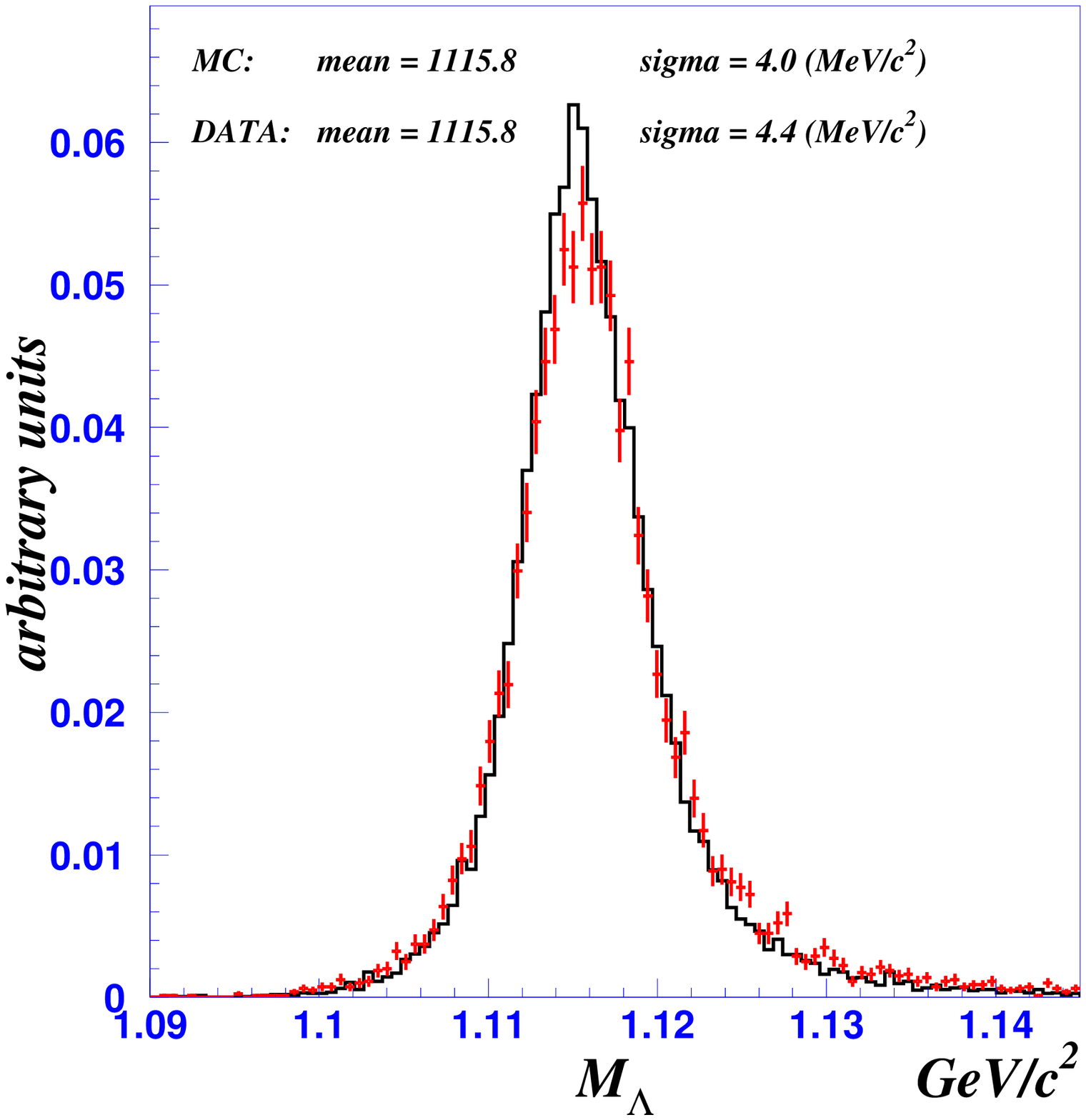,width=0.49\linewidth}}
\caption{
Invariant mass distributions for identified \ko 's ~(left) and \lam 's (right).}
\label{invariant_masses_k_l}
\end{figure}

\vo's allow to check
the quality of the drift chamber reconstruction by computing invariant masses
corresponding to different
neutral strange particle hypotheses (\ko, \lam, \alam):
$$M^2 = M^2_+ + M^2_- + 2 (E^+ E^- - P^+ P^- \cos \theta)$$
where $M_+$ and $M_-$ are the masses of positive and negative outgoing 
particles, $E^+$ and $E^-$ are their energies and $\theta$ is the angle between them.
Figure \ref{invariant_masses_k_l} shows \ko ~and \lam ~normalized invariant mass 
distributions for both data and Monte Carlo. 
The widths of these distributions are related to the momentum resolution and 
are in good agreement with what 
is expected from figure \ref{fig:dc_reso_real}.

\subsection{Test of the global alignment of the drift chambers} 

The alignment procedure described in section~\ref{sec:alignement} may 
lead to a systematic displacement of the calculated 
wire positions 
with respect to the real positions, so that a straight track
would appear to be slightly curved. This would obviously bias the momentum 
measurement. We show now how we used $V_0$ decays to measure such
a bias and eventually correct for it.


If we call the curvature bias 
$\varepsilon$ 
($\varepsilon = < 1/p_{rec}-1/p_{true}>$), 
its influence on the above mentioned reconstructed invariant mass 
can be written as: 
$$M \simeq M_T + \varepsilon \frac{\partial M}{\partial \varepsilon} = M_T + \frac{\varepsilon}{M} \left( \frac{1}{2} \frac{\partial M^2}{\partial \varepsilon} \right)$$
where $M$ stands for reconstructed and $M_T$ for 
the world average experimental value~\cite{PDG} 
of the invariant mass. The term  $\frac{1}{2} \frac{\partial M^2}{\partial \varepsilon}$ can be expressed as:

$$ 
 \frac{1}{2} \frac{\partial M^2}{\partial \varepsilon}=\left(P^+-P^-\right) \left(P^+P^- \cos \theta - E^+E^-+\frac{M_+^2M_-^2}{E^+E^-}\right)+\frac{P^+P^-}{E^+E^-}\left(P^-M_+^2-P^+M_-^2 \right)
$$

The evaluation of the momentum bias consists 
in fitting the distribution 
$M = f (\partial M / \partial \varepsilon)$ by a straight line. 
The fitted $M_T$ must be in good agreement with what is expected 
from~\cite{PDG}
and the momentum bias 
$\varepsilon$ is given by the slope parameter of the fit.

From all \vo ~decays reconstructed in the NOMAD setup, only the \ko's are symmetric 
in the decay phase space and have the same energy loss 
(dE/dx) for the daughter particles. 
The bias evaluation has been performed using a sample of 10540 \ko's 
with a purity of 99\%~\cite{Cyril}.

\begin{figure}[ht]
\centerline{\epsfig{figure=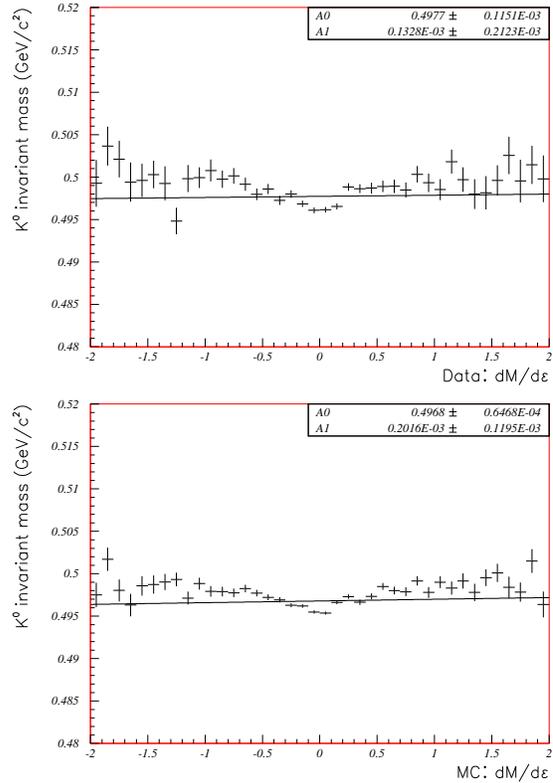,width=8cm}}
\caption{Invariant mass as a function of $\partial M/\partial \varepsilon$ for both data and Monte Carlo. The result of the fit is also shown. One notices
that the distribution significantly departs from a straight line, however
in the same way in the simulation and the data, and symmetrically around zero.
Since the measurement of the curvature bias only relies on the difference
between negative and positive sides of the distribution, we can measure the bias
using the procedure described in the text.
The systematic variations around the fit are
below 1\%, and we interpret them as small inaccuracies
in the track model used in the track fit, which show up when averaging 
over a large number of tracks.}
\label{fig:biais_courbure_by_ko}
\end{figure}

The obtained results (see figure \ref{fig:biais_courbure_by_ko}) 
are the following: in units $\rm (TeV/c)^{-1}$ $\varepsilon\ = \ 0.1 \pm 0.2$ 
in data and $\varepsilon\ = \ 0.2 \pm 0.1$ in simulated events. It is only
by chance that the value measured in real data comes out to be compatible with zero, and we did not have to correct the wire positions 
obtained by the alignment procedure (section~\ref{sec:alignement}).

The results show that there is no momentum bias.
One can give a physical interpretation of the found values.
A bias of 0.1 $\rm (TeV/c)^{-1}$ corresponds to a 10 $\rm TeV/c$ particle
reconstructed as a straight track (infinite momentum).
The bias on the momentum measurement of a 100 $\rm GeV/c$ track is of the order of 1\,\%
and it falls down to 0.1\,\% for a 10 $\rm GeV/c$ track (both values are below the intrinsic resolution
of the drift chambers).

The mean momentum of secondary particles 
produced in neutrino interactions 
in the NOMAD detector 
is lower than 10 $\rm GeV/c$.
We can consider that the momentum bias has a negligible effect on the estimation of track momenta.

%
%
\section{Conclusions}
\label{sec:conclusions}
The primary aim of the NOMAD experiment was to search for 
$\rm \nu_\mu \rightarrow \rm \nu_\tau$ oscillations, using
kinematical criteria to sign the presence of the $\tau$ production and decay.
This was made possible thanks to a set of large drift chambers
which at the same time 
provided
the target 
material
for neutrino interactions
and the tracker of charged particles.
The technology used to produce the drift field allowed
the high density of measurement points which would be difficult
to achieve by conventional techniques. Furthermore, these chambers were
built at a reasonable cost.
The chambers ran satisfactorily during 4 years, and the NOMAD experiment 
was able to push
by more than one order of magnitude the previous limits on  
$\rm \nu_\mu \rightarrow \rm \nu_\tau$~\cite{NOMAD_OSC} 
and $\rm \nu_e \rightarrow \rm \nu_\tau$~\cite{NOMAD_OSC1}
oscillation probability in 
a region of neutrino masses relevant for cosmology. 
The chambers played also a crucial role in
several precise studies of particle production in 
neutrino interactions~\cite{NOMAD_PHYS}.

%
\section*{Acknowledgements}
\label{sec:acknowledgements}
The Commissariat \`a l'Energie Atomique (CEA) and the Institut National de
Physique Nucl\'eaire et de Physique des Particules (IN2P3/CNRS) supported the 
construction of the NOMAD drift
chambers. 
We are grateful to the technical staff of these two institutions, namely
to the following departments:
SGPI (Service de Gestion des Programmes et d'Ing\'enierie), 
SED (Service d'Etudes des D\'etecteurs), 
SIG (Service d'Instrumentation G\'en\'erale), 
SEI (Service d'Electro-nique et d'Informatique).
We particularly acknowledge the help of D. Le Bihan for the relations with 
industry,
J.-L. Ritou for safety and quality insurance,
P. Nayman for his expertise on electromagnetical compatibility problems,
M. Serrano and R. Zitoun for their help in the software development 
and P. Wicht for the detector integration at CERN.\\
We would also like to warmly thank the whole NOMAD collaboration 
for their financial and technical support in solving the problems encountered
with the strip glueing.
Special thanks are due to 
L. Camilleri, L. Di Lella, M. Fraternalli, J.-M. Gaillard and A. Rubbia,
as well as J. Mulon, K. Bouniatov, I. Krassine and V. Serdiouk 
for their strong involvment
and to all the physicists and technicians who participated to the CERN repair 
workshop.

%
%

\end{document}